\documentclass[twocolumn,showpacs,preprintnumbers,amsmath,amssymb,floatfix]{revtex4}

\usepackage{graphicx}
\usepackage{dcolumn}
\usepackage{bm}
\usepackage{colordvi}
\usepackage{epsfig}
\usepackage{graphicx}
\usepackage{color}
\usepackage{booktabs}

\newcommand{\bp}{\boldsymbol{\partial}}

\newcommand{\degree}{\ensuremath{^\circ}}

\begin{document}

\title{Nonlinear response of dense colloidal suspensions under oscillatory shear: 
Mode-coupling theory and FT-rheology experiments}

\author{J.M.~Brader$^1$, M.~Siebenb{\"u}rger$^2$, M.~Ballauff$^2$
K.~Reinheimer$^3$, M.~Wilhelm$^3$, S.J.~Frey$^4$, F. Weysser$^5$ and M.~Fuchs$^5$ }

\affiliation{
$^1$Department of Physics, University of Fribourg, CH-1700 Fribourg, Switzerland}
\affiliation{ 
$^2$Helmholtz Zentrum f\"ur Materialien und Energie, D-14109 Berlin, Germany}
\affiliation{
$^3$Karlsruhe Institute of Technology, D-76128 Karlsruhe, Germany}
\affiliation{
$^4$Institut Charles Sadron, Universit\'e de Strasbourg, CNRS UPR 22, 
23 rue du Loess, 67034 Strasbourg, France}
\affiliation{
$^5$Fachbereich Physik, Universit\"at Konstanz, D-78457 Konstanz, Germany}


\begin{abstract}
Using a combination of theory, experiment and simulation we investigate the 
nonlinear response of dense colloidal suspensions to large amplitude oscillatory shear flow. 
The time-dependent stress response is calculated using a recently developed schematic 
mode-coupling-type theory describing colloidal suspensions under externally applied flow. 
For finite strain amplitudes the theory generates a nonlinear response, characterized by 
significant higher harmonic contributions. 
An important feature of the theory is the prediction of an ideal glass transition at 
sufficiently strong coupling, which is accompanied by the discontinuous appearance of a 
dynamic yield stress. 
For the oscillatory shear flow under consideration we find that the yield stress 
plays an important role in determining the non linearity of the time-dependent stress 
response. 
Our theoretical findings are strongly supported by both large amplitude oscillatory (LAOS)
experiments (with FT-rheology analysis) on suspensions of thermosensitive core-shell particles dispersed in water and 
Brownian dynamics simulations performed on a two-dimensional binary hard-disc mixture. 
In particular, theory predicts nontrivial values of the exponents governing the final 
decay of the storage and loss moduli as a function of strain amplitude which are in 
excellent agreement with both simulation and experiment. 
A consistent set of parameters in the presented schematic model achieves to jointly 
describe linear moduli, nonlinear flow curves and large amplitude oscillatory spectroscopy. 
\end{abstract}

\maketitle 

\section{Introduction}
A standard method to probe the viscoelastic character of a material is to measure 
the time dependent stress response to an externally 
applied oscillatory shear field \cite{larson1}. 
The simplicity of oscillatory shearing experiments presents distinct practical 
advantages when compared to other flow protocols and thus makes desirable a systematic 
method for the rheological characterization of a material on the basis of the periodic stress response alone. 
For small strain amplitudes the shear stress is a simple harmonic function, oscillating 
with the fundamental frequency dictated by the applied strain field. 
The details of the microscopic interactions underlying the macroscopic stress response 
are encoded in the familiar storage ($G'$) and loss ($G''$) moduli of linear response.
General aspects of the viscoelastic character of the material can thus be inferred 
from the magnitudes of the moduli as a function of frequency.

While, for many systems of interest, the linear response regime is well understood, 
for practical applications, 
such as the production and processing of materials in industry \cite{coussot}, 
it is necessary to consider deformations of finite, often large, amplitude.  
In the nonlinear regime, the stress response to a sinusoidal excitation 
contains higher harmonic contributions, which arise from the nonlinearity of the underlying 
constitutive relation expressing the stress as a functional of the strain  
\cite{larson2,wilhelm1,wilhelm2}. 
For many complex materials, consideration of the fundamental frequency alone  
proves insufficient for describing the physical mechanisms at work for finite strain 
amplitude.
Analysis based purely on the linear complex modulus as a function of frequency can thus be expected to give 
only a partial mechanical characterization  of the system under study 
(see e.g. \cite{mckinley,mckinley_preprint}). 
This failing is found to be particularly pronounced for yield stress materials 
such as aqueous foams \cite{fielding} and, as we will argue in the present work, 
colloidal suspensions close to, or beyond, the point of dynamical arrest. 
Although such systems are predominately elastic in character, they exhibit a complex 
transient response to oscillatory shear in which the viscous dissipation mechanism 
present at small strain amplitudes crosses over to a plastic flow as the amplitude is increased. 
The nonlinear stress response reflecting the onset of plastic flow gives rise to a 
strong increase in the amplitudes of the higher harmonics.

The emerging discipline of Fourier transform rheology (FT-rheology), originating 
in the work of Wilhelm and co-workers (see, e.g. \cite{wilhelm1,wilhelm2,Wilhelm07,wilhelm_99}), 
aims to quantify the nonlinear response of complex fluids by analyzing the harmonic 
structure of the stress signal measured in large amplitude oscillatory 
shear (LAOS) experiments (for recent developments see \cite{rogers}). 
Despite considerable progress on the experimental side, the theoretical 
description of the nonlinear regime remains unsatisfactory. 
Theoretical treatments capable of capturing higher harmonic contributions have  
been largely restricted to phenomenological models based on the ideas of continuum rheology 
\cite{larson2,mckinley_preprint,Wilhelm07,cho,hassager,noll}.
A more refined description of the nonlinear response is provided by mesoscopic models 
in which the time evolution of explicit coarse-grained degrees of freedom is governed by 
specified dynamical rules \cite{mes1,mes4,derec}. 
While such approaches are capable of capturing generic features of the response, 
they are not material specific and make no explicit reference to the underlying particle 
interactions.

Recently, progress in making the connection between microscopic and macroscopic 
levels of description has been made for the case of dense colloidal suspensions subject to 
time-dependent flow \cite{joeprl_07,joeprl_08}. 
The developments in classical nonequilibrium statistical mechanics presented in 
\cite{joeprl_07} and \cite{joeprl_08} extend earlier work focused on the simpler, 
but fundamental, case of steady shear flow 
\cite{fuchs_cates_PRL,fc_jrheol}. 
The mode-coupling-type approximations employed in 
\cite{joeprl_07,joeprl_08,fuchs_cates_PRL,fc_jrheol} capture the slow 
structural relaxation leading to dynamical arrest in strongly coupled systems 
(i.e. dispersions at high volume fraction or with a strongly attractive potential 
interaction), with the consequence that the macroscopic flow curves $\sigma(\dot\gamma)$ attain a 
finite value in the limit of vanishing rate, for states which would be glasses or 
gels in the absence of flow. 
The finite value of the stress in the slow flow limit identifies the {\em dynamic} 
yield stress. 
The relationship between the dynamic yield stress and its more familiar static counterpart 
is analogous to that between stick and slip friction in engineering applications. 
A prediction of particular importance made by the the MCT-based approaches of \cite{joeprl_07,joeprl_08,fuchs_cates_PRL,fc_jrheol} is that the dynamic 
yield stress appears discontinuously as a function of coupling strength, in clear contrast 
to mesoscopic models \cite{mes1,mes4} which predict a continuous power law dependence.  
The notion of yield stress was considered in a more general and abstract sense in \cite{pnas}, 
in which a dynamic yield stress surface, describing yielding under more general non-shear 
deformations, was calculated (see also \cite{joe_review}).

Although the closed, microscopic constitutive equation presented in \cite{joeprl_08} is of 
considerable generality, the combined difficulties of a large time-scale 
separation between microscopic and structural relaxation times, spatial anisotropy, and 
lack of time-translational invariance presented by many problems of interest makes 
direct numerical solution of the equations impossible at the present time. 
In order to both facilitate numerical calculations and expose more transparently the 
essential physics captured by the fully microscopic theory of \cite{joeprl_08} a 
simplified `schematic' model has been proposed \cite{pnas}. 
Schematic models have proved invaluable in the analysis and assessment of microscopic 
mode-coupling approaches, both for quiescent systems \cite{MCTequations} and under 
steady shear flow \cite{faraday}, in each case providing a simpler set of equations 
which aim to retain the essential mathematical structure of the fully microscopic 
theory. 
While the schematic model reduction performed in \cite{pnas} leads to loss of the 
`first principles' character of the approach, the mathematical connections 
between full and schematic theory nevertheless serve to elevate the schematic model 
above purely phenomenological approaches. 

In the present work we will consider application of the schematic model derived in 
\cite{pnas} to the problem of large amplitude oscillatory shear. 
Although the tensorial schematic model of \cite{pnas} is closely related to the earlier 
$F_{12}^{\dot\gamma}$ model derived in \cite{faraday}, application of the tensorial model 
to a simple shear flow geometry does not exactly reproduce the $F_{12}^{\dot\gamma}$ model. 
The study of time-dependent flows, not considered in earlier 
work, revealed that corrections to the original $F_{12}^{\dot\gamma}$ model were necessary 
to capture correctly the response to rapidly varying flows. 
The modifications thus introduced lead to small differences in the steady state rheological 
predictions. 
Nevertheless, the present schematic models describes the same phenomenology as the previous model 
\cite{hajnal_scaling} when applied to steady shear.

Comparison of theoretical predictions with experimental data for thermosensitive core-shell 
particles, dispersed in water, has been performed using the $F_{12}^{\dot\gamma}$ model \cite{faraday}. 
These particles have the very convenient feature that the volume fraction of the system may be 
varied continuously over a considerable range, simply by tuning the temperature of the system. Moreover, 
the finite polydispersity in particle size effectively suppresses crystallization, such that studies of dense fluid and glassy states are not complicated by an intervening fluid-crystal transition. 
In a series of works, theory and experiment have been compared for the flow curves under steady shear 
\cite{fuchs_ballauff,crassous0} and, more recently, for both flow curves and linear response moduli 
\cite{crassous,winter}. 
A particular strength  of the $F_{12}^{\dot\gamma}$ model (inherited by the more recent model of 
\cite{pnas}) is that both flow curves and linear viscoelastic moduli can be simultaneously and accurately fitted over many decades of shear rate and frequency, respectively, using a consistent and physically meaningful set of fit parameters.  
In \cite{winter} a combination of experimental techniques were employed, which enabled measurement of 
the flow curves and linear response moduli over eight and nine orders of magnitude in shear rate and 
frequency, respectively 
\cite{winter}. 
Although certain discrepancies between experiment and theory at low frequencies remain 
to be fully understood, the general level of agreement is impressive. 
Reassuringly for the schematic models, the complete microscopic MCT calculations possible for the linear 
response moduli agree with the data from the monodisperse samples on the $40$\% error level \cite{crassous0}. 

The nonlinear rheology of thermosensitive microgel particles (identical to those considered in the present 
work) was addressed in a recent experimental study, focused on the stress response to steady and large amplitude 
oscillatory flow \cite{carrier}. 
In addition to a study of the stress overshoot following the onset of shear flow (see also \cite{zausch}), both 
the strain dependence of the storage and loss moduli and the higher harmonic contributions were analyzed. 
Despite employing the same thermosensitive  particles and LAOS flow protocol, the study \cite{carrier} 
should be regarded as complementary to the present work. 
In \cite{carrier} volume fractions well above random close packing were 
investigated ($\phi>0.64$), suggesting considerable deformation of the particles themselves, whereas we focus 
here on packing fractions around the glass transition. Moreover, emphasis in the present work is placed on 
assessing the MCT based schematic theory presented in \cite{pnas} for a nontrivial flow history, namely 
large amplitude oscillatory shear, and comparison of the theoretical predictions with experiment. 
This comparison provides the first truly time-dependent test of this recently developed 
schematic model	beyond the simple case of step strain already considered in \cite{pnas}. 

The paper will be organized as 
follows: 
In Section \ref{fundamentals} we summarize the microscopic starting points underlying our theoretical 
approach, before proceeding to give a compact overview of the linear and nonlinear response of viscoelastic systems, 
relevant for the subsequent analysis. 
In Section \ref{theory_approach} we introduce the schematic MCT model and discuss its relation to 
previous work. 
In Section \ref{simulation} we discuss the Brownian dynamics simulation algorithm used to generate results 
supplementary to those of theory and experiment. 
Section \ref{experiment} contains the experimental details.  
In Section \ref{results} we first present purely theoretical results, in order to establish the phenomenology 
predicted by the schematic model. We then consider the results of our two-dimensional simulations before proceeding 
to analyze and fit the experimental data.  
Finally, in Section \ref{discussion} we discuss the significance 
of the present work and provide an outlook for future studies.

\section{Fundamentals}\label{fundamentals}

\subsection{Microscopic starting points}
The shear stress resulting from a general time-dependent shear strain of rate 
$\dot\gamma(t)$ is given by a generalized Green-Kubo relation \cite{joeprl_07,joeprl_08}
\begin{equation}
\sigma(t) = \int_{-\infty}^{\,t}\!\!dt'\; \dot\gamma(t')\,G(t,t').
\label{non-tti}
\end{equation}
Equation (\ref{non-tti}) is nonlinear in the shear rate due to the nonlinear functional 
dependence of the shear modulus $G(t,t')$ on $\dot\gamma(t)$. 
Within the microscopic framework developed in \cite{joeprl_07,joeprl_08} 
the modulus is identified as the correlation function of fluctuating 
stresses 
\begin{equation}
G(t,t')=\frac{1}{k_BTV}\langle\, \hat\sigma_{xy} e_-^{\int_{t'}^t ds\Omega^{\dagger}(s)}\,\hat\sigma_{xy}\,\rangle,
\label{exact_mod_nonlin}
\end{equation}
where $\hat\sigma_{\rm xy}\!\equiv\!-\sum_i\,  F_i^{\rm x}y_i$ is a fluctuating 
stress tensor element, formed by a weighted sum of the forces acting on the particles for 
a given configuration, $T$ is the temperature, $V$ is the system volume and 
$\langle\cdot\rangle$ indicates an equilibrium average. 
The particle dynamics to be considered in the present work are generated by the adjoint 
Smoluchowski operator \cite{dhont}
\begin{eqnarray}
\Omega^{\dagger}(t) = \sum_{i} D_0\,[\,\bp_i + \beta{\bf F}_i \,]\cdot\bp_i
+ D_0\dot\gamma(t) y_i \frac{\partial}{\partial x}
\label{adj_smol}
\end{eqnarray} 
where $\beta=1/k_BT$ and $D_0$ is the short time diffusion coefficient at infinite dilution. 
The time-ordered exponential function in Eq.(\ref{exact_mod_nonlin}) arises because $\Omega^{\dagger}(t)$ does 
not commute with itself for different times \cite{vankampen}. 
\par
An important approximation underlying Eq.(\ref{adj_smol}) (and thus (\ref{exact_mod_nonlin})) is that 
solvent induced hydrodynamic interactions (HI) between the colloidal particles are neglected. 
The diffusion coefficient entering Eq.(\ref{adj_smol}) is thus a scalar quantity and the external flow 
may be included using a prescribed (as opposed to self-consistently calculated) shear field 
$\dot\gamma(t)$.  
While the omission of HI may be inappropriate at high shear rates, for which 
hydrodynamically induced shear thickening can occur in certain systems, it is expected to represent 
a reasonable approximation for slowly sheared states close to the glass transition. 
Nevertheless, when attempting to fit experimental data using theoretical models based on 
Eq.(\ref{adj_smol}) it proves neccessary to include an empirical hydrodynamic correction accounting 
for the high frequency viscosity. 
In addition to the neglect of HI we make two, potentially more dangerous, assumptions: 
(i) $\dot\gamma(t)$ is taken to be spatially translationally invariant, which may become questionable when 
considering the flow response of dynamically arrested states. 
(ii) The shear field acts instantaneously. 
While this should be acceptable for certain flow histories the general status of this approximation is not clear.  

\subsection{Linear response}
Following standard convention, we consider an externally applied shear strain of the form 
\begin{equation}
\gamma(t)=\gamma_0\sin(\omega t).
\label{strain} 
\end{equation}
The time translational invariance of the shear field (\ref{strain}) gives rise to an explicit 
dependence of the modulus (\ref{exact_mod_nonlin}) upon two time arguments.  

For small deformation amplitudes ($\gamma_0\ll 1$) the strain dependence of the shear 
modulus may be neglected, such that Eq.(\ref{non-tti}) provides a linear relationship between 
$\dot\gamma(t)$ and $\sigma(t)$. 
This leads to the approximation
\begin{eqnarray}
G(t,t')=G_{\rm eq}(t-t'),
\label{Gtti} 
\end{eqnarray}
where $G_{\rm eq}(t)$ denotes the time translationally invariant equilibrium shear modulus. 
Substitution of (\ref{strain}) and (\ref{Gtti}) into 
Eq.(\ref{non-tti}) and employing trigonometric addition formulas leads directly to 
the familiar linear response result
\begin{equation}
\sigma(t)=\gamma_0G'(\omega)\sin(\omega t) + \gamma_0G''(\omega)\cos(\omega t),
\label{stress}
\end{equation}
where $G'(\omega)$ and $G''(\omega)$ are the storage and loss moduli, 
respectively, defined by 
\begin{eqnarray}
G'(\omega)&=&\omega\!\int_0^{\infty}\!\!dt' \sin(\omega t')\,G_{\rm eq}(t') 
\label{gp}
\\
G''(\omega)&=&\omega\!\int_0^{\infty}\!\!dt' \cos(\omega t')\,G_{\rm eq}(t'). 
\label{gpp}
\end{eqnarray}
Furthermore, Eq.(\ref{stress}) can be rewritten as 
\begin{eqnarray}
\sigma(t) = \gamma_0|G(\omega)|\sin\left(\omega t + \delta(\omega)\right),
\label{stress_phase}
\end{eqnarray}
where the complex modulus is given by $G=G'+ iG''$ and the phase shift by 
$\delta=\arctan(G''/G')$. If $G''(\omega)=0$ the response is purely elastic, 
in phase with $\gamma(t)$ ($\delta=0$). 
In the case $G'(\omega)=0$ dissipation dominates and the response is in 
phase with $\dot\gamma(t)$ ($\delta=90\degree$). 

\subsection{Nonlinear response}\label{nonlinear}
It should be clear at this stage that the familiar linear response form (\ref{stress}) 
is a direct consequence of the convolution integral which results from inserting 
the time translationally invariant equilibrium function (\ref{Gtti}) into 
Eq.(\ref{exact_mod_nonlin}). For finite strain amplitudes, the dependence of the modulus 
upon two time arguments prevents the simple trigonometric manipulations leading to 
Eq.(\ref{stress}). Nevertheless, the non-sinusoidal stress response, $\sigma(t)$, is 
periodic with period $2\pi/\omega$, and may therefore be expressed as a Fourier series
\begin{eqnarray}
\sigma(t)= \gamma_0\sum_{n=1}^{\infty} G'_n(\omega)\sin(n \omega t) + 
\gamma_0\sum_{n=0}^{\infty} G''_n(\omega)\cos(n \omega t),
\notag\\
\label{fourier_series}
\end{eqnarray} 
where $G'_n$ and $G''_n$ are frequency dependent Fourier coefficients given by \cite{boas}
\begin{eqnarray}
G'_n(\omega)=\frac{\omega}{\pi}\int_{-\pi/\omega}^{\pi/\omega}\!dt\;\sigma(t)\sin(n\omega t)
\label{gpn}\\
G''_n(\omega)=\frac{\omega}{\pi}\int_{-\pi/\omega}^{\pi/\omega}\!dt\;\sigma(t)\cos(n\omega t).
\label{gppn}
\end{eqnarray}
In the limit $\gamma_0\rightarrow 0$ the coefficients $G'_1$ and $G''_1$ reduce to the familiar 
linear response moduli. 
It should be noted that we retain the $n=0$ term in the second sum of (\ref{fourier_series}) 
in order to leave open the possibility of a stress offset.

Employing manipulations analogous to those leading from (\ref{stress}) to (\ref{stress_phase}) the Fourier 
series (\ref{fourier_series}) may be expressed in the form
\begin{eqnarray}
\sigma(t)= \gamma_0\sum_{n=1}^{\infty} I_n(\omega)\sin(n \omega t + \delta_n(\omega)),  
\label{fourier_series_phase}
\end{eqnarray} 
where the amplitude is given by $I_n=|G'_n+iG''_n|$ and the phase shifts by $\delta_n(\omega)=\arctan(G_n''/G_n')$. 
In analyzing our theoretical, experimental and simulation results we will focus on 
the behaviour of both the generalized moduli $G'_n$ and $G''_n$ and the amplitude and 
phase shift, $I_n$ and $\delta_n$, of the fundamental ($n=1$) and higher harmonics ($n>1$) 
as a function of  the control parameters.   

Following a period of transient response after initiation of the strain field 
(switching on the rheometer) the system enters a stationary state, demonstrating a periodic 
stress response. 
Although, to some extent, an issue of semantics, it is important that the `stationary' 
state presently under consideration be distinguished from `steady' states, of the kind 
achievable by application of a time-independent shear flow. 
The stationary state is simply a well characterized and periodic transient and is thus 
influenced by additional physical mechanisms (e.g. thixotropy) which are irrelevant for steady 
states.
In a physical system the stationary response must be independent of the 
direction of shear, leading to a stress $\sigma(t)$ symmetric in $\dot\gamma(t)$. 
The mirror symmetry of the constitutive equation has the consequence that only odd terms 
contribute to the Fourier series (\ref{fourier_series_phase}). 
The appearance of even harmonics in the analysis of experimental data is often an indication 
of boundary effects, such as wall slip, or other inhomogeneities of the flow \cite{Wilhelm07}.

\par
Important physical interpretation may be given to the coefficient $G''_1$ by considering 
the energy dissipated per unit volume of material per oscillation cycle 
\begin{eqnarray}
E_d = \int_{-\pi/\omega}^{\pi/\omega}\!dt\;\sigma(t)\dot\gamma(t).
\label{dissipation}
\end{eqnarray}
Substitution of the strain field (\ref{strain}) and the Fourier series 
(\ref{fourier_series}) into Eq.(\ref{dissipation}) leads to 
\begin{eqnarray}
E_d = \gamma_0^2\pi\, G''_1(\omega),
\label{dissipation1}
\end{eqnarray}
(see also \cite{hyun_wilhelm}).
Thus, for a sinusoidal strain of the form (\ref{strain}), energy is dissipated only at 
the input frequency. 
The coefficient $G''_1$ therefore has the same interpretation in the non-linear regime as 
in the linear regime: it determines the dissipation of energy over an oscillation cycle. 
The remaining coefficients in the series, $G'_n$ and $G''_{n>1}$, thus collectively describe 
the reversible storage and recovery of elastic energy. 

\subsection{Lissajous plots}
A standard way to graphically represent the relationship between $\gamma(t)$ and 
$\sigma(t)$ is via the Lissajous representation, in which trajectories are shown 
in the $\gamma^*,\sigma^*$ plane, where $\gamma^*=\gamma/\gamma_{\rm max}$ 
and $\sigma^*=\sigma/\sigma_{\rm max}$ are the strain and stress, normalized by their 
maximum values \cite{lissajous_old}. 
In this representation, a general {\em linear} viscoelastic response is characterized by an 
ellipse, symmetric about the line $\gamma^*=\sigma^*$, point symmetric with respect to the 
origin plus two mirror planes. 
The two limiting cases of a 
purely elastic and a purely dissipative response are thus characterized by a line 
and a circle, respectively. 
In the nonlinear regime considerable deviations from ellipticity are observed.  
The specific character of these deviations can indicate whether a material is, for example, 
strain hardening or strain softening (an increase/decrease of $G'$ with strain amplitude), and thus provides a useful, albeit qualitative, 
'rheological fingerprint` of a given material \cite{mckinley,mckinley_preprint}. 
For a general nonlinear response, the area enclosed within the closed loop trajectory of 
a Lissajous figure is directly related to the dissipated energy via the integral in 
Eq.(\ref{dissipation}). 
This lends an appealing physical interpretation to the Lissajous representation and 
provides a direct visual impression of the dissipative character of the response.

\section{Theoretical approach}\label{theory_approach}
\subsection{Schematic  model}
As noted in the introduction, the approximate microscopic constitutive theory developed in 
\cite{joeprl_07,joeprl_08} enables first-principles prediction of the rheological behaviour 
of dense colloidal dispersions.  
However, the simultaneous occurrence of spatial anisotropy and non-time translational 
invariance hinders numerical solution of the equations when addressing concrete problems. 
The schematic model presented in \cite{pnas} provides a simplified set of equations which, 
it is hoped, capture the essential physics contained within the full equations, while remaining 
tractable for numerical implementation.  


Within the schematic reduction, the modulus is expressed in terms of a single-mode transient density correlator
\begin{eqnarray}\label{modulus}
G(t,t')=v_{\sigma}\Phi^2(t,t')
\end{eqnarray}
where $v_{\sigma}$ is a parameter measuring the strength 
of stress fluctuations. 
The approximation underlying (\ref{modulus}) is that stress fluctuations relax as a result of 
relaxations in the density (viz. structural relaxation). 
The assumption that $v_{\sigma}$ is independent of strain is a simplifying assumption 
which could be relaxed if neccessary. 
The microscopic theory of \cite{joeprl_08} predicts both the temporal and wavevector dependence 
of the transient density correlater under applied flow. 
The schematic, single mode, density correlator (normalized to $\Phi(t,t)=1$) represents, in some 
non-specific sense, a `typical' correlator of the microscopic theory. 
It is obtained from solution of a nonlinear integro-differential equation 
\begin{eqnarray}\label{schematic_eom}
\dot\Phi(t,t') 
+ \Gamma\bigg(
\Phi(t,t') + \int_{t'}^{t}ds\, m(t,s,t')
\dot\Phi(s,t')
\bigg) =0.
\notag\\
\end{eqnarray}
The single decay rate $\Gamma$ sets the time-scale and would, within a microscopically based theory, 
depend upon both structural and hydrodynamic correlations. 
The overdots in Eq.(\ref{schematic_eom}) imply differentiation with respect to the first time argument.
The memory function $m(t,s,t')$ appears in Eq.(\ref{schematic_eom}) as a generalized friction kernel, which can be formally identified as the correlation function of certain stress fluctuations. 
Making the assumption that these stress fluctuations may be expressed in terms of density fluctuations 
(both become slow close to the glass transition) leads to a tractable expression for 
$m(t,s,t')$ as a quadratic functional of the transient density correlator and, thus, a closed theory.  
A somewhat surprising consequence of the formal calculations presented in \cite{joeprl_07,joeprl_08} 
is that the memory function possesses three time arguments. The presence of a third time argument, which 
would have been difficult to anticipate on the basis of quiescent MCT intuition, has important consequences for rapidly varying flows (e.g. step strain \cite{joeprl_07}). 
Within the schematic model the memory function is given by
\begin{eqnarray}\label{schematic_memory}
m(t,s,t')=h(t,t')\,h(t,s)\,[\,\nu_1\Phi(t,s)+\nu_2\Phi^2(t,s)\,].
\notag\\
\end{eqnarray}  
Following conventional MCT practice the parameters $\nu_1$ and $\nu_2$ represent, in an unspecified 
way, the role of the static structure factor in the microscopic theory and are chosen as 
$v_2=2$ and $v_1=2(\sqrt{2} -1) + \epsilon/(\sqrt{2}-1)$. The separation parameter 
$\epsilon$ is a crucial parameter within our approach and encodes the thermodynamic statepoint 
of the system by measuring the distance from the glass transition. 
Negative values of $\epsilon$ correspond to fluid states and positive values to 
glass states. 
Setting $h$ equal to unity in Eq.(\ref{schematic_memory}) recovers the well known  $F_{12}$ model, originally introduced by G\"otze \cite{MCTequations,goetze_zeit,sjoegren}.
The linear term in $\Phi$ which appears in Eq.(\ref{schematic_memory}) is absent from the microscopic 
mode-coupling expression, but turns out to be necessary for a faithful reproduction of its asymptotic 
properties within a single-mode theory. 
Under simple  shear flow, the $h$-functions in the memory kernel (\ref{schematic_memory}) serve to accelerate the loss of memory caused by the affine advection of density fluctuations. 
The assumption that the same function $h$ may be used to incorporate both the strain accumulated between 
$t$ and $t'$ as well as that between $t$ and $s$ is an approximation, made to keep the theory as simple as possible. 
Taking account of the required invariance with respect to flow direction suggests the simple ansatz
\begin{eqnarray}
h(t,t')=\frac{\gamma_c^2}{\gamma_c^2 +  \gamma^2},
\label{hfunc}
\end{eqnarray}
where $\gamma\equiv\gamma(t,t')=\int_{t'}^{t}ds\,\dot\gamma(s)$ and the parameter $\gamma_c$ 
sets the scale of strain.

Eq.(\ref{non-tti}) and (\ref{modulus})-(\ref{hfunc}) provide a closed constitutive theory 
which depends upon three adjustable parameters $(v_{\sigma},\Gamma,\gamma_c)$ and two control 
parameters $(\epsilon,\dot\gamma)$ representing the coupling strength and applied shear rate. 
As the schematic model under discussion is implicitly based on the Smoluchowski dynamics 
described by Eq.(\ref{adj_smol}), the influence of HI is neglected. 
While this is not important for capturing correctly the qualitative features of the 
rheological response, quantitative comparison requires a simple hydrodynamic correction at 
high frequencies. 
The simplest approximation, which we will employ 
in the present work, is to empirically add an extra term to the shear modulus 
\begin{eqnarray}\label{hydro_correction}
G(t,t') \rightarrow G(t,t') + \eta_{\infty}\,\delta(t-t').
\end{eqnarray}
The high frequency viscosity, $\eta_{\infty}$, is thus introduced into the model, 
describing the viscous processes which occur on timescales much shorter than the structural 
relaxation time. 
The correction (\ref{hydro_correction}) has the consequence that the stress acquires an 
extra term, linear in $\dot\gamma$, and the Fourier coefficient $G''_{1}$ is shifted by a term 
linear in $\omega$.

\subsection{Strain-rate frequency superposition}
An alternative mode-coupling-type approach, describing the collective density fluctuations 
of dense colloidal fluids under shear, is provided by the work of Miyazaki {\em et al}. \cite{miyazaki1,miyazaki2,miyazaki3}.
By considering time-dependent fluctuations about the steady state a closed (scalar) constitutive 
equation has been derived and applied to colloidal dispersions in two-dimensions under 
steady shear \cite{miyazaki1,miyazaki2} and in three dimensions (subject to additional isotropic 
approximations) under large amplitude oscillatory shear \cite{miyazaki3}. 
Given the very different nature of the approximations underlying the present MCT-based theory  
\cite{joeprl_07,joeprl_08,fuchs_cates_PRL,fc_jrheol} and that of 
\cite{miyazaki1,miyazaki2,miyazaki3} (fluctuating hydrodynamics vs. projection operator methods) 
it is interesting that the final expressions (e.g. the memory function vertices entering the equation 
of motion for the transient correlator) are rather similar, at least for the special case of steady 
shear. 
For the case of large amplitude oscillatory shear, however, the theory presented in \cite{miyazaki3} 
differs clearly and fundamentally from the microscopic approaches to time-dependent shear developed in \cite{joeprl_07,joeprl_08} and, consequently, from the schematic model of \cite{pnas} to be 
employed in the present work. 
The theoretical developments of Miyazaki {\em et al}. \cite{miyazaki3} motivated the authors to 
propose the principle of `strain-rate frequency superposition' as a probe of structural relaxation in soft materials \cite{wyss}.

The essence of the Miyazaki {\em et al}. approach can be captured by a simple schematic model, which we will 
elaborate upon below. 
In \cite{miyazaki3} the authors took the theory which they had developed for steady shear flow 
\cite{miyazaki1,miyazaki2} and replaced the steady shear rate $\dot\gamma$ appearing in the equation 
of motion for the correlator, by the time-dependent shear rate $\dot\gamma(t)=\gamma_0\omega\cos(\omega t)$,
describing oscillatory flow. 
This rather {\em ad hoc} treatment gives rise to equations with a mathematical structure appropriate 
for steady flows and ignores the more realistic, although more complicated, history dependence of 
theories developed to treat non-steady flows specifically \cite{joeprl_07,joeprl_08}. 
On the basis of the results obtained for the strain amplitude dependence of the storage and loss moduli 
(notated as $G'_{1}, G''_{1}$ in the present work) it was argued that the time-dependence of the strain-rate 
field $\dot\gamma(t)=\gamma_0\omega\cos(\omega t)$ is not essential for understanding the viscoelastic response, and that it is sufficient to consider the strain-rate amplitude $\gamma_0\omega$ alone. 
The relevant timescale is thus identified as $(\gamma_0\omega)^{-1}$, rather than $\omega^{-1}$. 
Within the context of schematic mode-coupling equations, this assumption may be expressed by the following 
memory function
\begin{eqnarray}\label{mr_memory}
m(t)=\frac{[\,\nu_1\Phi(t)+\nu_2\Phi^2(t)\,]}{1 + (\gamma_0\omega t)^2},
\notag\\
\end{eqnarray} 
which, together with the equation of motion
\begin{eqnarray}\label{mr_eom}
\dot\Phi(t) 
+ \Gamma\bigg(
\Phi(t) + \int_{0}^{t}dt'\, m(t-t')
\dot\Phi(t')
\bigg) =0,
\notag\\
\end{eqnarray}
the shear modulus
\begin{eqnarray}\label{mr_modulus}
G(t,t')=v_{\sigma}\Phi^2(t-t'),
\end{eqnarray}
and Eq.(\ref{non-tti}) provides a closed theory for $\sigma(t)$. 
In fact, Eqs.(\ref{mr_memory})-(\ref{mr_modulus}) are identical to the $F_{12}^{\dot\gamma}$ 
model \cite{MCTequations,faraday}, with a steady shear-rate $\dot\gamma=\gamma_0\omega$. 
An important consequence of assuming the dominance of the timescale $(\gamma_0\omega)^{-1}$ is that 
all states, even those which would be glasses in the absence of flow, become fluidized by an 
applied oscillatory shear field, regardless of the amplitude $\gamma_0$. 
Whether or not a vanishingly small value of $\gamma_0$ is really sufficient to restore ergodicity 
to dynamically arrested states is unclear and presents a fundamental question, with important implications 
for the existence of a linear response regime. 

Despite capturing approximately the amplitude dependence of $G'_{1}, G''_{1}$,
describing the response at the fundamental frequency, higher harmonics are ignored in the approach 
of \cite{miyazaki3}. The absence of higher harmonic contributions within the theory of Miyazaki 
{\em et al}. can be traced back to the assumption that the time-dependence of $\dot\gamma(t)$ is 
irrelevant and that this can be represented by the constant $\gamma_0\omega$. 
Within the present context this has the consequence that the memory function (\ref{mr_memory}) and 
correlator, given by solution of (\ref{mr_eom}), are constrained to be time-translationally invariant 
(viz. depend on a single correlation time only). 
While this assumption is clearly at odds with the underlying variations in the strain field 
(for which a dependence on the `waiting time' is to be expected),  
it nevertheless serves to capture first-order corrections to linear 
response theory, while remaining relatively easy to implement numerically. 

The theory developed in \cite{miyazaki3} is quasi-linear, in the sense that $\sigma(t)$ remains 
a simple sinusoid, but with an amplitude and phase shift which depend non-linearly on 
$\gamma_0$. 
Attempts to justify the neglect of higher harmonics have been based on the 
fact that the ratio of the third harmonic amplitude to that of the fundamental 
remains smaller than approximately $20$\%, for a wide range of systems \cite{miyazaki3,wyss}.
However, in order to draw a fair conclusion, it is important to consider 
the sum $\sum_{n>1} I_n(\omega)/I_1$, rather than $I_3/I_1$ alone, when assessing the physical 
relevance of higher harmonic contributions. 
Various experimental studies on colloidal dispersions (see e.g. \cite{carrier}) show clearly 
that the higher harmonics can collectively account for up to half of the total signal, which is 
not a small effect. 
This observation serves to emphasize the importance of truly nonlinear theories, which confront 
directly the non-time translational invariance of the correlation functions, thus going beyond the convolution approximation to Eq.(\ref{non-tti}). 

\section{Computer simulation}\label{simulation}
To provide a point of reference for our theoretical calculations we have 
performed two-dimensional simulations on a system hard discs 
undergoing Brownian motion in an external 
shear field. 
The simulations are designed to solve approximately the many-body problem of 
a system of interacting Brownian particles under shear flow. 
The same Smoluchowski dynamics \cite{dhont} underlies the microscopic mode coupling theories 
of \cite{joeprl_07} and \cite{joeprl_08} which form the basis of the schematic model employed in the 
present work \cite{pnas}. 
We choose to simulate a two-dimensional system for two reasons: 
(i) The computational resources required are significantly reduced with respect to simulation 
of three-dimensional systems and thus enables improved statistics to be obtained.  
(ii) Recent microscopic studies of the quiescent mode-coupling theory in two-dimensions have 
revealed behaviour broadly similar to that found in three dimensional calculations \cite{bayer}. 
We thus expect the reduced dimensionality of our simulation system to be of little 
consequence for qualitative comparison with the present theory and experimental data.
\par
The basic concept of the algorithm has been described in detail in three dimensions in \cite{scala} and its adaptation to two dimensions can be found in \cite{henrich}.
We consider a binary mixture of hard discs with the diameters of $D_s=1.0$ and $D_b=1.4$  with equal particle number concentrations and a total amount of $N=N_s+N_b=1000$ hard discs in a two-dimensional simulation box of volume $V$ with periodic Lees Edwards boundary conditions. The total two-dimensional volume fraction is then given by $\phi_{tot} =  \frac{N \pi}{4V} (D_b^2 + D_s^2)$. We employ this system in order to suppress crystallisation effects. 
The mass $m$ of the particles and $k_B T$ is set equal to unity. We choose our coordinate axes such that flow is in the $x$-direction and the shear gradient is in the $y$-direction.
The Brownian timestep was chosen to $\delta t = 0.01$ as in \cite{henrich}. This results in a short time diffusion constant of $D_0=0.05$. 
To implement a time-dependent, oscillatory shear rate, at each Brownian timestep the shear rate is set to 
its new value
\begin{equation}
\dot\gamma(\tau_B)= \gamma_{0}\,\omega \cos(\omega \tau_B),
\end{equation}
and all particle velocities are freshly drawn from the Gaussian distribution 
with $\langle {\bm v}^2\rangle=2$ and 
$\langle {\bm v} \rangle=\dot{\gamma}(\tau_B) y(\tau_B)$.
Between two Brownian timesteps the shear rate is kept constant. 
The strain $\gamma(t)$ can therefore be obtained using
\begin{equation}
\gamma(t) = \sum_{\tau_B \in \left[ 0,t \right]} \dot{\gamma}(t_B)\delta t\,,
\end{equation}
which leads to $\gamma(t) = \int_{0}^{t} {\rm d}t \; \dot\gamma(t) $ in the limit 
of $\delta t \to 0$.
At every Brownian timestep the part $\dot{\gamma}(\tau_B) y(\tau_B)$ guarantees a linear 
velocity profile as a linear shear flow is imposed on every particle, depending on 
its $y$-position. For all simulations the frequency was set to $\omega = 0.001$ which leads to Pe$_\omega\equiv\omega D^2/4D_0=0.05$.
\par
The average quantity of interest in the present work is the time dependent potential part of the shear stress $\sigma_{xy}(t)= \frac{1}{V} \left \langle \sum ({\bm F}(t)_{ij})_x ({\bm r}(t)_{ij})_y \right \rangle$, with the relative force components of particle $i$ and $j$ $({\bm F}(t)_{ij})_x$ and the particles relative distance 
component $({\bm r}(t)_{ij})_y $ for a given time $t$.
As we consider hard particles the forces must be calculated from the collision events. 
By observing the collisions within a certain time window $\Delta \tau_c =  \left[ t_k, t_k+\Delta \tau_c \right]$  
for a given time $t_k$, forces may be extracted using the change of momentum which occurs during the 
observation time. 
This leads to the evaluation algorithm for the stress at time $t_k$
\begin{equation}
 \sigma(t_k) = \left \langle  \frac{1}{\Delta \tau_c} 
 \sum_{t_c \in \left[ t_k, t_k+\Delta \tau_c \right]} 
 (\Delta {\bm v}(t_c)_{ij})_x ({\bm r}(t_c)_{ij})_y \right \rangle_s,
\notag\\
\end{equation}
where summation is over all collisions after time $t_c$ within the time window 
$\Delta \tau_c$. 
The procedure effectively sums the momentum changes $(\Delta {\bm v}_{ij})_x$ in $x-$ 
direction multiplied by the relative distance of the particles $({\bm r}_{ij})_y$ 
in $y-$ direction. The brackets $\langle ... \rangle_s$ denote the different 
simulation runs.\\
At a total volume fraction of $\phi_{tot} = 0.81$ which is slightly above the glass transition for this system (estimated to be at $\varphi = 0.79$ on the basis of simulated flow curves \cite{henrich}) we prepared $4000$ independent sets for each amplitude $\gamma_0 \in \lbrace 0.001, 0.003, 0.009, 0.01, ...,0.09, 0.1, 0.2, 1.0, 10.0, 100.0 \rbrace$
As the system starts from a non stationary state it is necessary to wait for the system to reach it's long time 
asymptote (which we found to be the case after undergoing two full oscillations) before meaningful averages can be taken.\\

\section{Experiment}\label{experiment}
\subsection{Characterization of the latex particles}
The polydisperse latex particles consist of a solid core of poly(styrene) (PS) onto which a thermosentitive 
network of crosslinked poly(N-isopropylacrylamide) (PNIPAM) is affixed \cite{winter}. 
The degree of crosslinking of the shell due to the crosslinker N,N'-methylenebisacrylamide is 
2.5 Mol \%. As exactly the same particles were used for this work as in \cite{winter}, the latices have a temperature dependent size (hydrodynamic radius in nm $R_H$ = -0.7796 $\cdot$T +102.4096 with T the temperature 
in $^{o}$C below 25$^{o}$C) and a polydispersity of 17\% \cite{winter}. 
All experiments were done in an aqueous solution of 0.05M KCl to screen residual charges 
which emerge from the synthesis of the particles. The solid content of the suspension was determined by 
comparing the weight before and after drying and was found to be 8.35 $wt$\%. 
Neither the MCR 301 measurements nor the FT-rheology measurements at 15.1$^\circ$C lead to a 
significant change of the solid content (+0.02 $wt$\%).
However, the FT-rheology measurements at the remaining two temperatures (18.4$^\circ$C and 20$^\circ$C) had a slightly different solid content (9.02 $wt$\%) due to the physical relocation of 
the rheometer to another laboratory and some additional solvent evaporation. 
The effective volume fraction $\phi_{\rm eff}$ was calculated by using the correlation of 
mass concentration $c$, hydrodynamic radius and effective volume fraction found in the inset of 
fig. 6 in \cite{winter}, which is given by $c \cdot R_H^3 = 9.67 \cdot 10^{-17}\,g \cdot \phi_{\rm eff}$ for the different temperatures. 
For the temperature of 15.1$^\circ$C a volume fraction $\phi_{\rm eff}$ of 0.65, for 18.4$^\circ$C 
a $\phi_{\rm eff}$ of 0.60 and for 20.0$^\circ$C a $\phi_{\rm eff}$ of 0.57 was found. 
In previous work \cite{winter} the glass transition for this system was found to be at  $\phi_{\rm eff}=0.64$. 
Given a polydispersity of $17$\% the theory of Sch\"artl and Sillescu \cite{sillescu} predicts random close packing 
at $\phi_{\rm eff}=0.68$.

\subsection{Rheological Experiments} 
The rheological experiments cover the range from the linear to the strongly nonlinear regime. The rheological measurements were performed with two instruments a MCR 301 from Anton Paar with a Cone Plate geometry (diameter: 50 mm, cone angle: 0.991$\degree$) and the FT-rheological measurements with an ARES rheometer (Rheometrics Scientific) with a Cone Plate geometry (diameter: 50 mm, cone angle: 0.04 rad). All measurements were performed after a preshear of 100s$^{-1}$ lasting 200s and a waiting time of 10s. For the rheological measurements, a solvent trap (for the ARES instrument equipped with a sponge drawn with water) is used to prevent evaporation. A thin paraffin-layer covers additionally the solution to prevent an exchange with the atmosphere above.\\
With the MCR 301 the measurements at 15, 18 and 20$^\circ$C were done. The flowcurves were measured from 5$\cdot$10$^{-4}$ to 1000s$^{-1}$ with a time ramp of 2500s to 20s and then a waiting time of 10s followed by a flowcurve measured from 1000 to 5$\cdot$10$^{-4}$s$^{-1}$ with a time ramp of 20 to 2500s. For 15$^\circ$C flowcurves in the shear rate range of 1$\cdot$10$^{-4}$s$^{-1}$ to 1000s$^{-1}$ with a timeramp of 10000s to 20s were perfomed. The frequency tests were performed at a deformation of 1\% starting from 10Hz to 0.001Hz with a time ramp of 20s to 1000s. All deformation tests were performed with a measurement time for each point of 100s.\\
For the FT-rheology oscillatory time sweep measurements were performed at the ARES instrument at frequencies of 1, 0.1 and 0.01Hz at different deformation amplitudes. The FT-signal were always recorded after some oscillations, so that the suspension reached the oscillatory stationary state. The first measurements were done at 15.1$^\circ$C and the frequency tests of the ARES and the MCR 301 coincided. For the two higher temperatures the measurements had to be performed for the same ARES instrument at a different room. Here an evaporation effect was observed, so that the temperature had to be adjusted to archieve the same behaviour in the frequency test. Instead of 18$\degree C$ a temperature of 18.4$\degree C$ and instead of 20$\degree C$ a temperature of 20.9$\degree C$ was used. Typically for the nonlinear FT-rheology measurements with the time sweep tests 40 oscillations for 1 Hz excitation were applied, whereas 10 oscillations for 0.1 Hz and 9 oscillations for 0.01 Hz. To obtain a FT-spectrum from the raw time data we performed a discrete, complex, half-sided fast Fourier transformation \cite{bracewell,wilhelm2,wilhelm_99}.

\begin{figure}
\includegraphics[width=7cm]{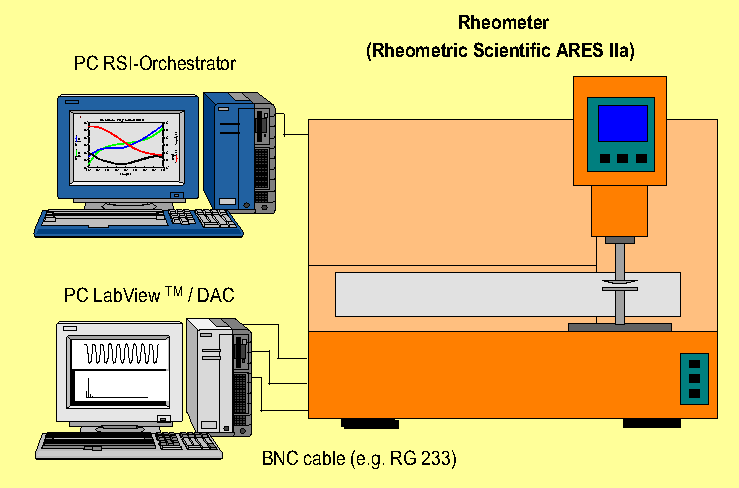}
\caption{{\color{red}(Experiment)} 
The experimental set-up of FT-rheology with two independent 
controlling computers, one for normal rheology and one for FT-rheology 
\cite{wilhelm2}.
}
\label{exp_f1}
\end{figure}

 For further information of the setup, the measuring principal and FT-analysis we would like to refer to \cite{
klein,neidhoefer}

\subsection{Fourier Transform Rheology Apparatus}
The FT-rheological setup consists of a Rheometrics Scientific advanced rheometer expansion 
systems (ARES) and a computer which either controls the rheometer via a serial cable as well 
as detects the strain and torque force outputs via BNC cables (see Fig.\ref{exp_f1}). 
The ARES rheometer is a 
strain-controlled rheometer equipped with a dual range force rebalance transducer (100 FRT) 
capable of measuring torques ranging from 0.004 mNm to 10 mNm, specified by the manufacturer. 
It has a high resolution motor, applying frequencies from $10-5$ rad/s to $500$ rad/s and deformation 
amplitudes ranging from $0.005$ to $500$ mrad. A water bath adjusts the temperature range from 
$5^{\circ}$C to $95^{\circ}$C. 
The analog raw data of the measurements are digitized with a $16$-bit ADC. This ADC card has a 
maximum sampling rate of $50$ kHz per channel. 
Due to the high sampling rate the time between consecutive 
data points is very small compared to the timescale of rheological experiments. 
The loss of information by sampling the torque transducer data is negligible \cite{schmidt}. 
To achieve best results with respect to the signal-to-noise ratio, oversampling is applied. 
The ADC-card acquires the time data at the highest possible sampling rate and then preaverages 
them on the fly to reduce random noise. With this method the noise is reduced by a factor of 3 to 5 which could only be achieved by averaging multiple 
measurements \cite{dusschoten}. Within the set-up a 16-bit ADC card is implemented, which 
is able to discriminate steps.  The quantification resolution of the ADC card limits the ratio. 
It determines the minimum detectable intensity of weak signals 
by its ability to discriminate the intensity of the signal. The higher the bit number, the 
smaller the detectable intensity variation \cite{skoog}. After acquisition and digitization 
of the time data, they are handled with Matlab software \cite{klein,neidhoefer}.



\section{Results}\label{results}

\subsection{Theoretical predictions}

\subsubsection{Flow curves}
\begin{figure}[!t]
\includegraphics[width=8.5cm]{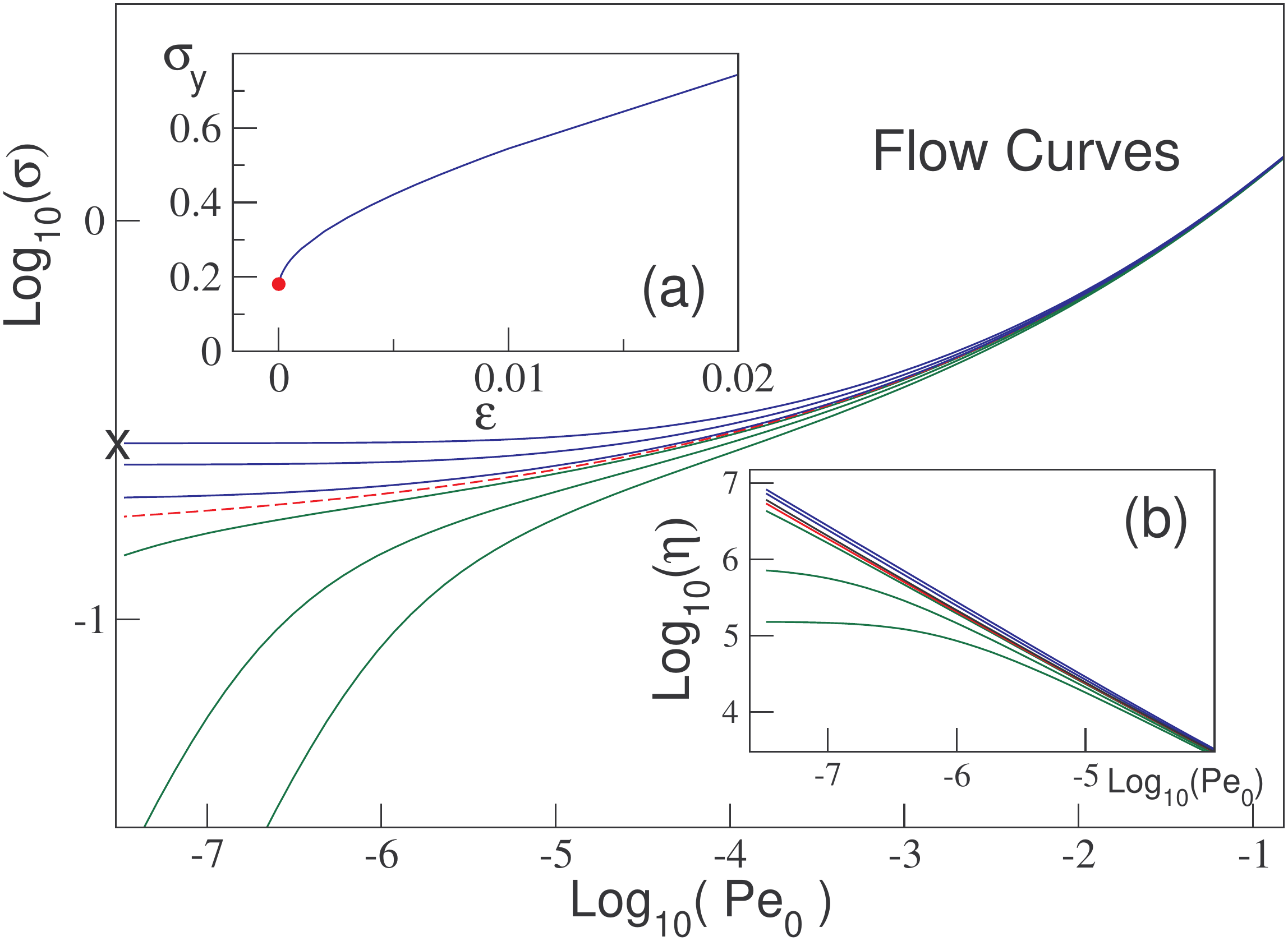}
\caption{{\color{red}(Theory)} 
The flow curves for fluid states $\epsilon= -10^{-3}, 
-5\times10^{-4}, -10^{-4}$ (green), at the critical point 
$\epsilon=0$ (red) 
and for glassy states $\epsilon=10^{-4}, 5\times10^{-4}, 10^{-3}$ (blue). 
In the glass ($\epsilon>0$) there exists a finite stress in 
the limit of vanishing shear rate, identified with the dynamical 
yield stress ($\lim_{\dot\gamma\rightarrow 0}=\sigma_y$). 
The cross indicates the yield stress value used in Figs.\ref{th_f3} and \ref{th_f4}.
Inset (a) shows the discontinuous emergence of a dynamical yield stress as a function 
of $\epsilon$. 
Inset (b) shows the viscosity $\eta=\sigma/\dot\gamma$. 
Calculations were performed with $\eta_{\infty}=0$. 
}
\label{th_f1}
\end{figure}

For given values of the parameters ($v_{\sigma},\gamma_c,\Gamma,\epsilon$) the schematic 
theory defined by (\ref{non-tti}) and Eqs.(\ref{modulus})-(\ref{hfunc}) enables prediction of the 
flow curve expressing the steady shear stress $\sigma$ as a function of shear-rate $\dot\gamma$. 
Fig.\ref{th_f1} shows a set of typical flow curves generated by the schematic MCT model 
for three fluid states ($\epsilon<0$), the critical state ($\epsilon=0$), and three states 
in the glass ($\epsilon>0$). 
The parameters employed for the theoretical calculations presented in Fig.\ref{th_f1} (as well as 
for Figs.\ref{th_f2}-\ref{th_f7}) are $v_{\sigma}=100$, $\gamma_c=0.15$, $\Gamma=100$ and $\epsilon=10^{-3}$. 
Experience with fitting the experimental data, to be considered in section \ref{exp_results}, shows that  
these choices represent sensible physical values for the model parameters. 
In the fluid, there exists a linear (Newtonian) regime for small shear-rates ($Pe_0\equiv\dot\gamma/\Gamma\ll 1$), for 
which the standard $F_{12}$ model result for the shear viscosity holds, 
$\sigma=\dot\gamma\int_0^{\infty}\!dt\, \Phi_{\rm eq}^2(t)\equiv \dot\gamma\eta$. 
Increasing the separation parameter to less negative values (corresponding to, e.g. an increase 
in the volume fraction) gives rise to an increase in $\eta$, reflecting the slowing of the 
structural relaxation time $\tau$, which dominates all transport properties within our MCT 
approach. 
For $\dot\gamma\tau>1$ the effect of shear starts to dominate the structural relaxation and 
the stress increases sub-linearly as a function of shear-rate, corresponding to shear thinning 
of the viscosity $\eta(\dot\gamma)=\sigma(\dot\gamma)/\dot\gamma$. 
At high shear-rates $\dot\gamma\tau\gg 1$ the present model yields $\sigma=\dot\gamma/\Gamma$ 
and needs to be supplemented by corrections which account for the high shear limiting 
viscosity (and which, in the absence of HI, aredetermined by the solvent contribution $\eta_{\infty}$). 

As $\epsilon\rightarrow 0^-$ the regime of linear response shifts to increasingly 
lower values of the shear-rate and disappears entirely at the (ideal) glass transition, 
$\epsilon=0$. 
For states in the glass there exists a finite stress in the limit of vanishing shear rate, 
identified as the dynamical yield stress 
($\lim_{\dot\gamma\rightarrow 0}\sigma(\dot\gamma)=\sigma_y$). 
Within idealized MCT based treatments the dynamical yield stress emerges discontinuously 
as $\epsilon$ is varied accross the glass transition (shown in the inset of Fig.\ref{th_f1}). 
It should be mentioned that the flow curves shown in Fig.\ref{th_f1} differ quantitatively 
from those of the extensively studied $F_{12}^{\dot\gamma}$ model \cite{faraday}, due to the inclusion of 
an additional prefactor $h(t,t')$ in the expression for the memory function 
(\ref{schematic_memory}). Nevertheless, the qualitative predictions of the theory for the flow 
curves are in full agreement with those of the $F_{12}^{\dot\gamma}$ model. 

\subsubsection{Linear response moduli}
\begin{figure}[!t]
\includegraphics[width=8.5cm]{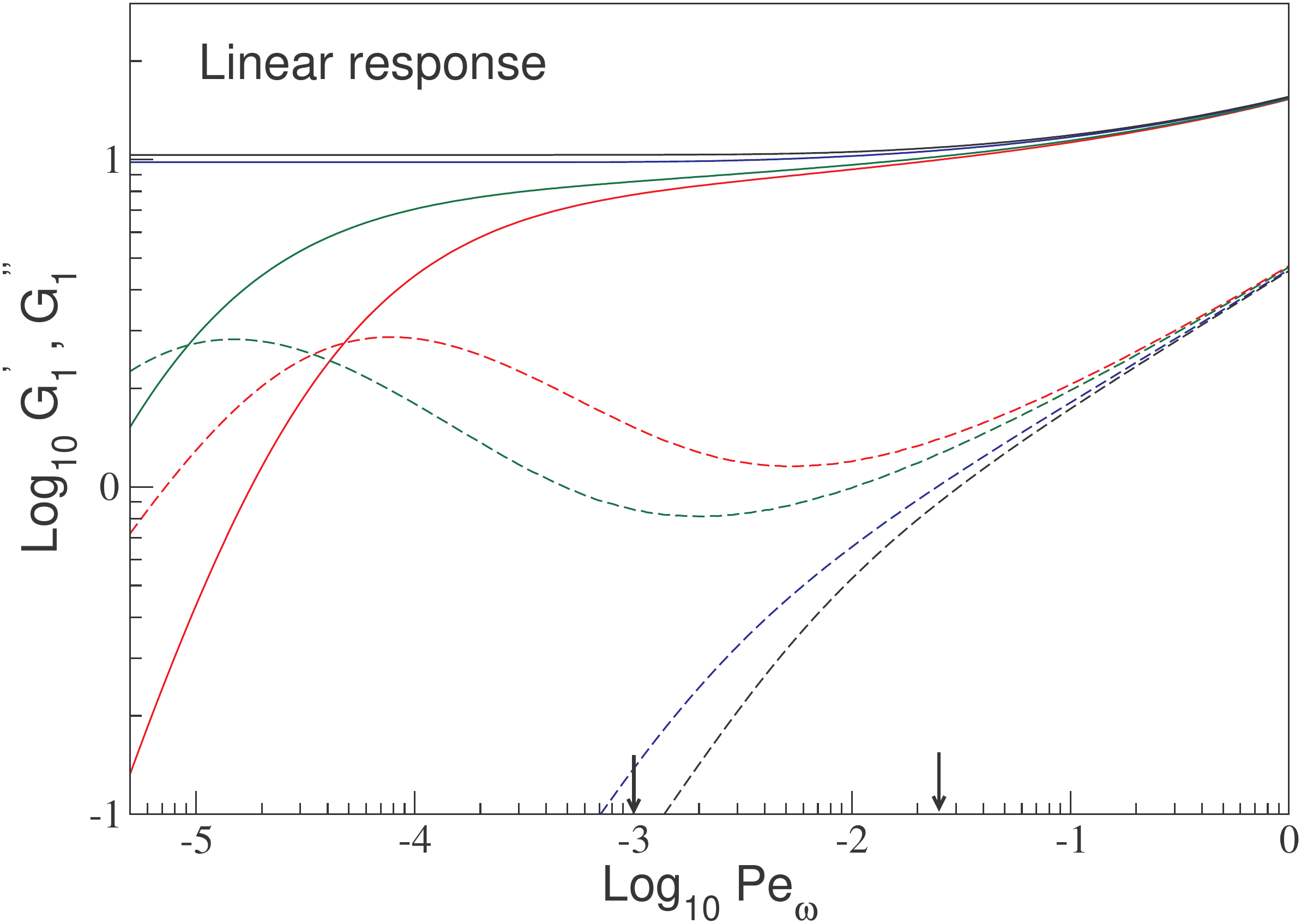}
\caption{{\color{red}(Theory)} 
The linear response moduli $G_1'$ (full lines) and $G_1''$ (broken lines) as a function 
of $Pe_{\omega}$ 
for two statepoints in the fluid $\epsilon=-0.001$ (red), $-0.0005$ (green) and two in 
the glass $\epsilon=0.0005$ (blue), $0.001$ (black). 
For fluid states the finite value of the structural relaxation 
timescale is reflected in the maximum in $G''$ and the consequent crossing of $G'$ and 
$G''$ at low frequency. The results presented here omit solvent 
hydrodynamics which may become relevant at $Pe_{\omega}\approx1$ (calculations were performed with $\eta_{\infty}=0$). 
These will be included 
in a simple approximate fashion when fitting the experimental data.  
The two arrows indicate the values $Pe_{\omega}=0.001$ and $0.025$ used to generate Figs.\ref{th_f3} and 
\ref{th_f4}, respectively.    
}
\label{th_f2}
\end{figure}

The linear storage and loss moduli, given by Eqs.(\ref{gp}) and (\ref{gpp}), respectively, 
are shown in Fig.\ref{th_f2} as a function of $Pe_{\omega}\equiv\omega/\Gamma$ for two fluid states ($\epsilon<0$) 
and two glassy states ($\epsilon>0$). 
In the fluid, the finite value of the structural relaxation timescale $\tau$ is reflected in the 
maximum of $G''$ and the crossing of $G'$ and $G''$ at low frequency. 
The fact that $G'$ remains notably larger than $G''$ at high frequencies is simply a result of 
neglecting the high frequency limiting viscosity $\eta_{\infty}$ in presenting our 
theoretical predictions. Setting $\eta_{\infty}=0$ in presenting the theory highlights the 
contribution of structural processes to the viscoelasticity.
In the glass, $G''$ goes to zero at low frequencies ($G''\!\sim\!\omega$) and $G'$ attains a finite 
low frequency value, identifying the transverse elastic constant $G_{\infty}$. 
Within our MCT approach, the elastic constant appears discontinously upon crossing 
the glass transition (i.e. $\lim_{\omega\rightarrow 0}G'(\omega)$ jumps from zero for 
$\epsilon\rightarrow 0^-$ to a finite value for $\epsilon\rightarrow 0^+$) thus demonstrating 
that the MCT indeed describes a transition to an amorphous solid. 

\subsubsection{Nonlinear stress response}
By numerical solution of the equation of motion (\ref{schematic_eom}) we obtain the 
non-time translational invariant density correlator $\Phi(t,t')$ and, via Eqs.(\ref{non-tti}) 
and (\ref{modulus}), the nonlinear stress response. 
The numerical algorithm requires the equation of motion to be discretized over the entire 
two-dimensional $(t,t')$-plane. 
While considerations of both causality ($t>t'$) and the periodicity of the correlator with respect to a 
translation in time ($\Phi(t+t_0,t'+t_0)=\Phi(t,t')$, where $t_0=\pi/\omega$) enable certain simplifications to be made, calculation of the 
correlator over many decades in time remains a computationally demanding task. 

\begin{figure}[!t]
\vspace*{0.1cm}
\includegraphics[width=9cm]{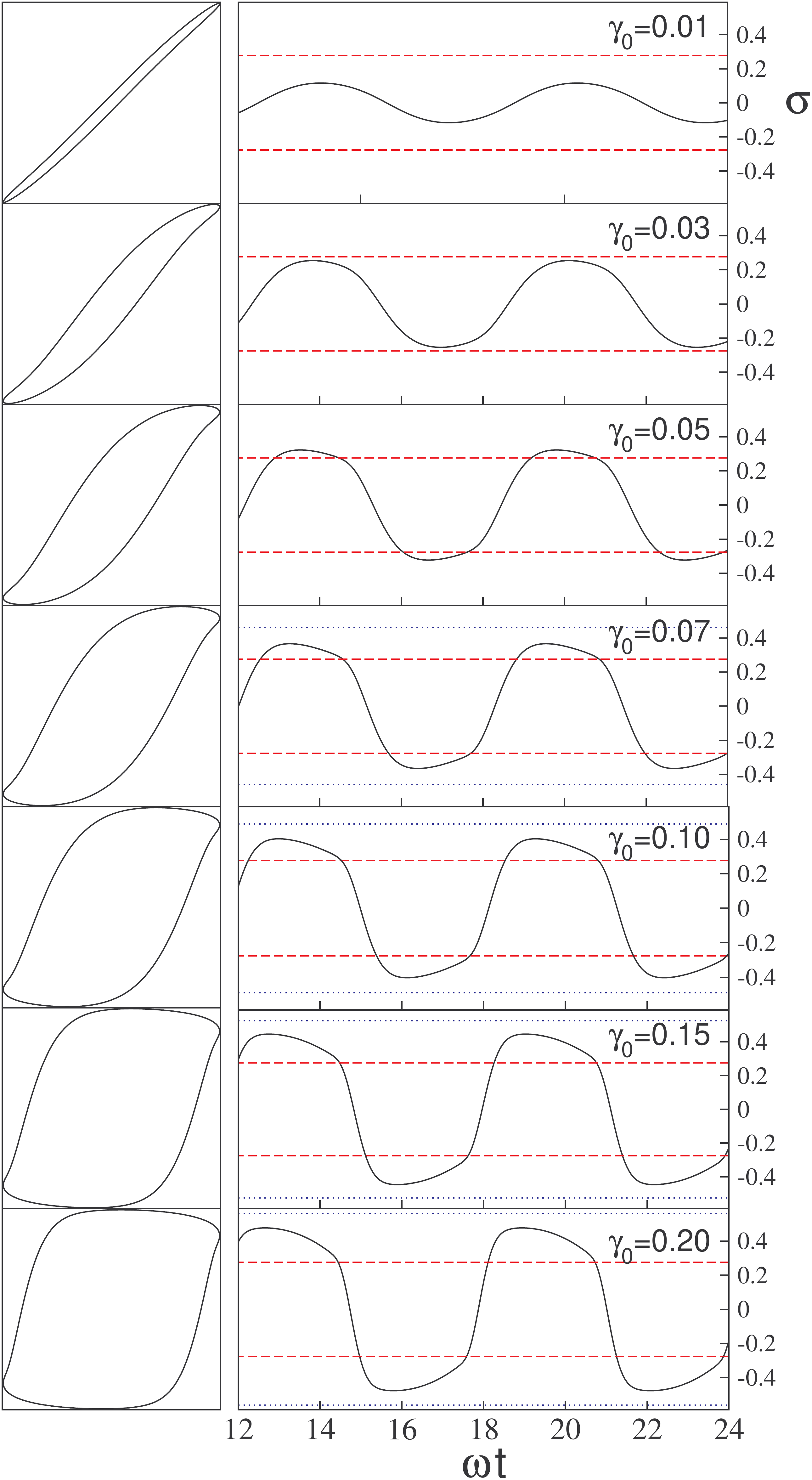}
\caption{{\color{red}(Theory)} 
The stress response of a glassy state to oscillatory strain 
calculated from our MCT based theory for strain amplitudes from 
$\gamma_0=0.01, 0.03, 0.05, 0.07, 0.10, 0.15$ and $0.20$.   
The associated Lissajous figures illustrate the nonlinear character of the response. 
The increase in dissipation with increasing $\gamma_0$ is reflected by the 
increasing area enclosed by the Lissajous curves. 
All calculations were performed at $Pe_{\omega}=0.025$ and $\epsilon=0.001$. 
The red horizontal broken lines indicate the dynamic yield stress obtained 
from the flow curve in Figure.\ref{th_f1} for $\epsilon=0.001$ 
($\sigma_y=0.2763$). 
The response becomes clearly nonlinear when the maximum of $\sigma(t)$ approaches 
the dynamical yield stress.  
The blue horizonal dotted lines provide an upper bound for the maximum of the time-dependent 
stress and are taken from the corresponding flowcurve in Fig.\ref{th_f1}.
}
\label{th_f3}
\end{figure}
 
Typical examples of the response $\sigma(t)$ for a glassy state ($\epsilon=0.001$) are shown in Fig.\ref{th_f3} for various values of the strain amplitude $\gamma_0$ at a fixed frequency. 
Alongside each of the time series we show the corresponding Lissajous curves indicating the extent 
of the dissipation (via the area enclosed, see Eq.(\ref{dissipation})) and the deviations from nonlinearity 
(discernable from the non-ellipticity of the loop). 
The value of $\epsilon$ employed to generate this figure generates a dynamic yield stress 
$\sigma_y=0.2763$, which is indicated in each panel of Fig.\ref{th_f3} by a broken red line. 
This can also be read-off from the appropriate flow curve in Fig.\ref{th_f1}.

For small amplitudes $\gamma_0\le 0.01$ the system is almost linear and responds in a 
predominately elastic fashion at the considered frequency. 
As $\gamma_0$ is increased, clear deviations from a sinusoidal response are apparent and higher harmonics 
start to contribute to the signal. 
In this nonlinear regime the stress response exhibits a characteristically flattened peak, with 
an asymmetry which increases as a function of $\gamma_0$. 
It is clear from the figure that the higher harmonics first become significant when the maximum value of the 
stress $\sigma(t)$ approaches the dynamic yield stress. 
An {\em idealized} yield stress material, subject to large amplitude oscillatory shear 
of vanishing frequency, would be expected to show a steady increase of the stress up to the yield point, 
beyond which the system begins to flow, maintaining $\sigma(t)\!=\!\sigma_y$ until the reversal of strain 
enables relaxation back to zero.  
If this were the case, then $\sigma(t)$ would be represented by a `clipped' signal, symmetric 
during loading and unloading of the sample. 

\begin{figure}[!t]
\vspace*{0.1cm}
\includegraphics[width=9cm]{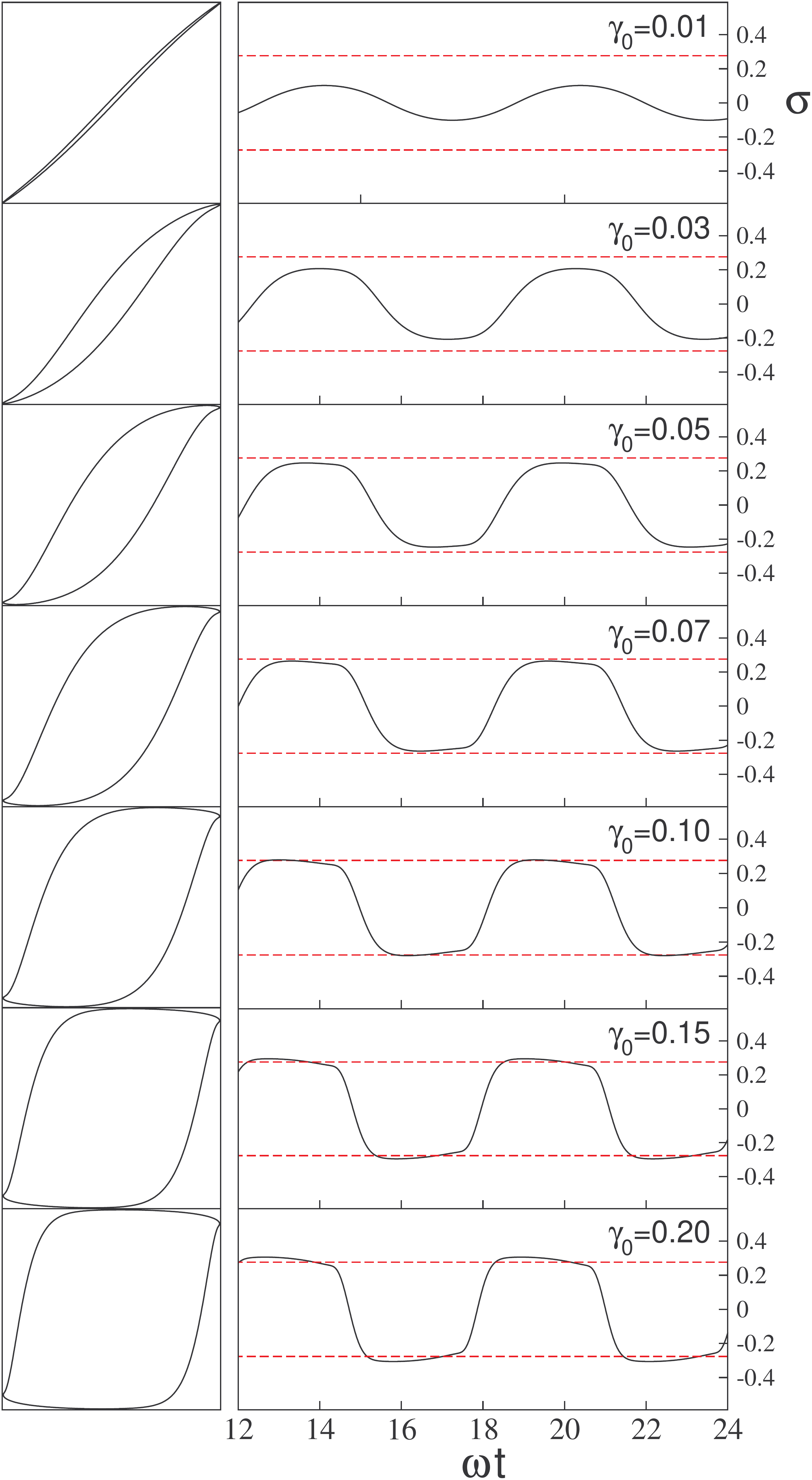}
\caption{{\color{red}(Theory)} 
As in Figure.\ref{th_f2} ($\epsilon=0.001$) but for a lower frequency 
$Pe_{\omega}=0.001$. At this value of $Pe_{\omega}$ the system is almost perfectly 
elastic in the linear regime ($G'\gg G''$ for $\gamma_0\ll 1$, see Figure.\ref{th_f4}). 
As the time-dependent stress exceeds the dynamical yield stress the signal becomes clipped. 
At this frequency $\sigma_{\rm max}$ lies very close to $\sigma_y$ and has thus been omitted 
for clarity. 
}
\label{th_f4}
\end{figure}

The results presented in Fig.\ref{th_f3}, however, have been generated for a low, but not vanishingly small, frequency. At finite frequency the maximum stress attainable in a system under steady shear is given by 
$\sigma_{\rm max}\equiv\sigma(\gamma_0\omega)$, where $\sigma(\dot\gamma)$ is the stress on the steady shear flow curve. 
This maximal stress under flow, $\sigma_{\rm max}$, is indicated by a blue dotted line in 
Fig.\ref{th_f3}, for the four largest strain amplitudes considered. 
In each case, once the stress exceeds the yield point the curve flattens, exhibiting a maximum which remains 
bounded from above by $\sigma_{\rm max}$. 
In the low frequency limit 
the lower and upper bounds to the peak value of $\sigma(t)$ become equal, 
$\sigma_{\rm max}\!=\!\sigma_y$, such that the signal becomes clipped at the yield point. 
In order to test this hypothesis further, we show in Fig.\ref{th_f4} stress responses generated using the same parameter set as employed in Fig.\ref{th_f3}, but for a frequency one order of magnitude lower. 
At this reduced frequency, the clipping of $\sigma(t)$ at yield is quite clear, although the the peak stress still slightly exceeds $\sigma_y$, due to the fact that the values of the 
strain-rate amplitude are not 
sufficiently small that $\sigma_{\rm max}$ has saturated to $\sigma_y$. 

\begin{figure}
\includegraphics[width=8.cm]{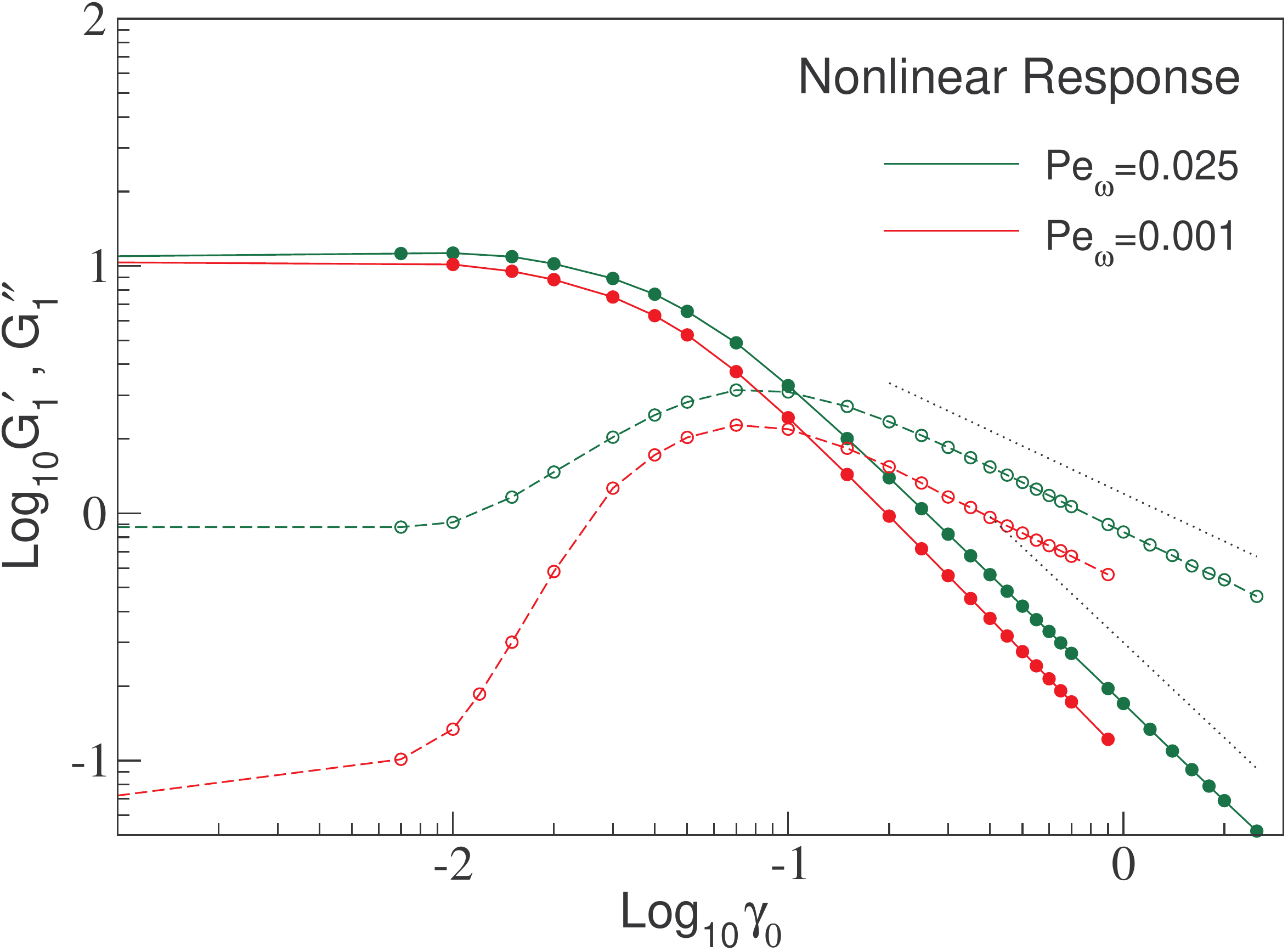}
\caption{{\color{red}(Theory)} 
The theoretical $G_1'$ and $G_1''$ as a function of strain for 
$\epsilon=0.001$ at frequencies $Pe_{\omega}=0.025$ and $Pe_{\omega}=0.001$. 
Points are the numerically calculated data points, lines are a guide for the eye. 
For large values of the strain the numerical data are well fitted by the 
power laws $G''\sim \gamma_0^{-\nu}$ and $G'\sim\gamma_0^{-2\nu}$ 
with $\nu=0.65$ (indicated by dotted lines). 
}
\label{th_f5}
\end{figure}

Two additional comments are in order regarding the results shown in Figs.\ref{th_f3} and \ref{th_f4}. 
Firstly, for low frequencies (e.g. the data shown in Fig.\ref{th_f4}) the full time-dependent stress signal can be rather faithfully reproduced by the simple approximation 
\begin{eqnarray}\label{simple_approx}
\sigma(t) = 
\begin{cases}
\;\gamma_0 [\, G_1'(\omega)\sin(\omega t) + G_1''(\omega)\cos(\omega t) \,] \;\;\sigma\le\sigma_y
\\
\;\;0 \hspace*{5.18cm}\sigma>\sigma_y
\end{cases}\notag\\
\end{eqnarray} 
where $G_1'$ and $G_1''$ are the lowest order coefficients in the Fourier series (\ref{fourier_series}). 
The naive picture sketched above is thus not completely correct. 
In order to describe correctly the sub-yield response, $\sigma(t)\le\sigma_y$, it is neccessary to 
incorporate the $\gamma_0$ dependence of the lowest order coefficients; linear response is insufficient.
It is also noteworthy that it is the {\em dynamic}, yield stress, which 
plays the crucial role in determining the time-dependent $\sigma(t)$. 
While the importance of dynamic yield in determining the oscillatory response is clear within the present approach, it remains to be seen whether this is a constraint introduced by employing a prescribed strain or, more significantly, an indication that the dynamic and static yield stresses are identical within our approximate theory. 
The simple approximation (\ref{simple_approx}) contains higher harmonic contributions as a result of the yield stress clipping criterion. 

\begin{figure}[!t]
\includegraphics[width=9cm]{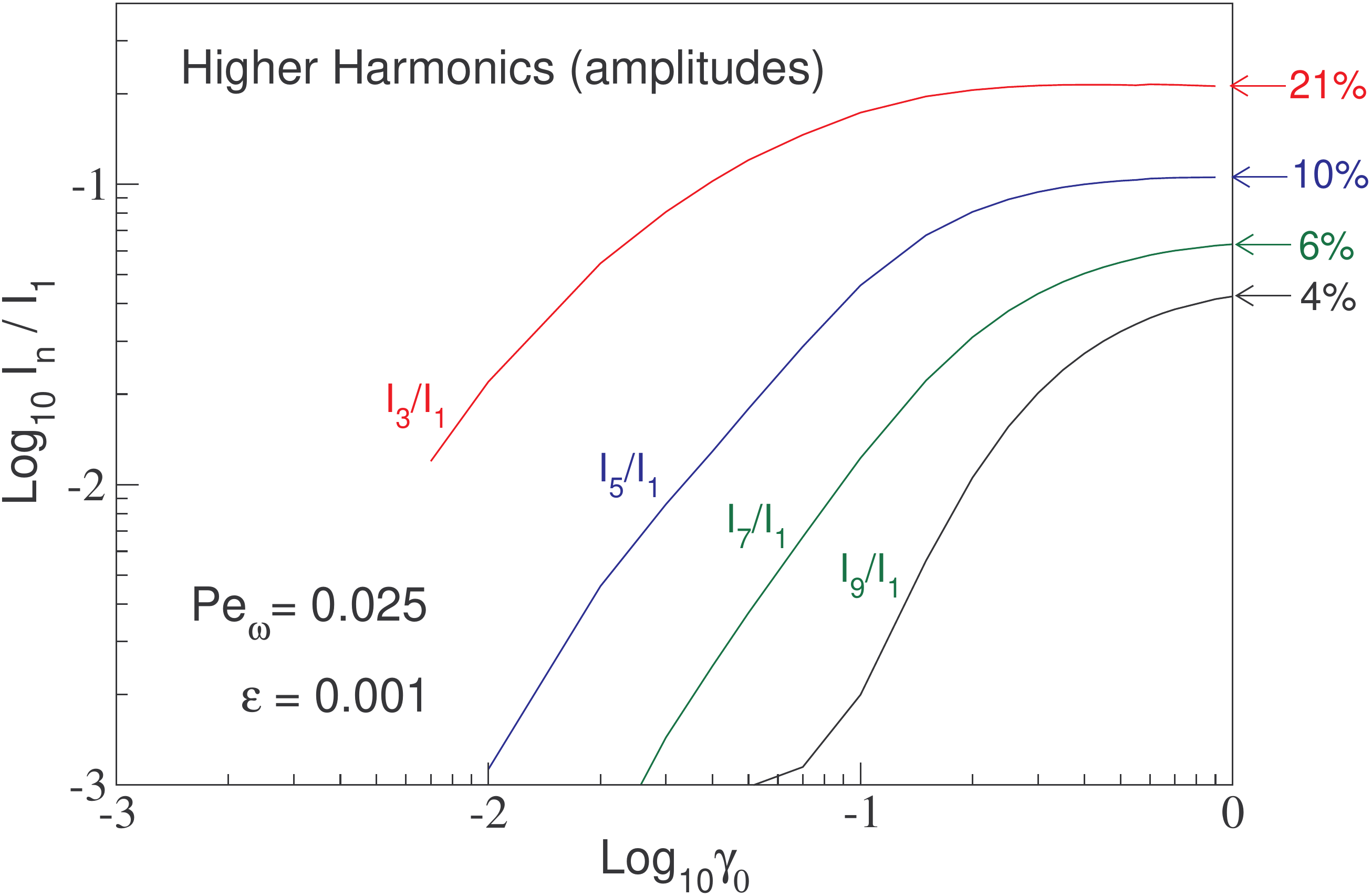}
\caption{{\color{red}(Theory)} 
The theoretical intensities of the third, fifth, seventh and ninth harmonic 
which contribute to the nonlinear stress response shown in Figure.\ref{th_f2}. 
At strains approaching unity the collective contribution of the higher harmonics 
accounts for approximately $50\%$  of the total $\sigma(t)$ signal.
($\epsilon=10^{-3}$ and $Pe_{\omega}=0.025$). 
}
\label{th_f6}
\end{figure}

\subsubsection{Fourier analysis}
In order to provide a more systematic analysis of the time signal $\sigma(t)$, we now consider its 
decomposition into Fourier modes and investigate the behaviour of the coefficients entering the series (\ref{fourier_series}) and (\ref{fourier_series_phase}) as a function of $\gamma_0$. 
We address first the strain amplitude dependence of $G'_1$ and $G''_1$, thus mimicing the ubiquitous 
`strain sweep' experiments generally used to assess the nonlinear response of a given material. 
In Fig.\ref{th_f5} we show typical results for the lowest order coefficients as a function of strain, for two different values of the excitation frequency.

For small values of $\gamma_0$ linear response is recovered and the values of $G'_1$ and $G''_1$ may be read-off 
from the data shown in Fig.\ref{th_f2}. 
The linear response regime persists up to around $\gamma_0=0.01$, beyond which $G''_1$ begins to increase gradually, 
reaching a maximum value at around $\gamma_0=0.1$. 
As noted in subsection \ref{nonlinear}, the coefficient $G'_1$ is proportional to the amount of energy dissipated per oscillation cycle. 
The increase in dissipation observed over the range $\gamma_0=0.01\rightarrow 0.1$ is probably connected to the increasing disruption of the microscopic `cage' structure of dense glassy systems, induced by the externally applied strain field. 
However, such microscopic interpretations remain purely speculative within the present context of schematic model calculations, for which there is no explicit spatial resolution of correlated density fluctuations.

Deeper insight into the microscopic mechanisms underlying the observed macroscopic response would be 
provided by solution of the full equations presented in \cite{joeprl_08}. 
The increase in $G''_1$ is associated with a decrease in $G'_1$ as a function of $\gamma_0$. 
In contrast to $G''_1$, there exists no simple physical interpretation of the coefficient $G'_1$ in the 
nonlinear regime. 
For strain amplitudes exceeding $\gamma_0=0.01$ the recoverable elastic energy becomes distributed over 
$G'_1$ and the higher harmonics, such that $G'_1$ loses its special status as a `storage modulus'. 

\begin{figure}[!t]
\includegraphics[width=8.8cm]{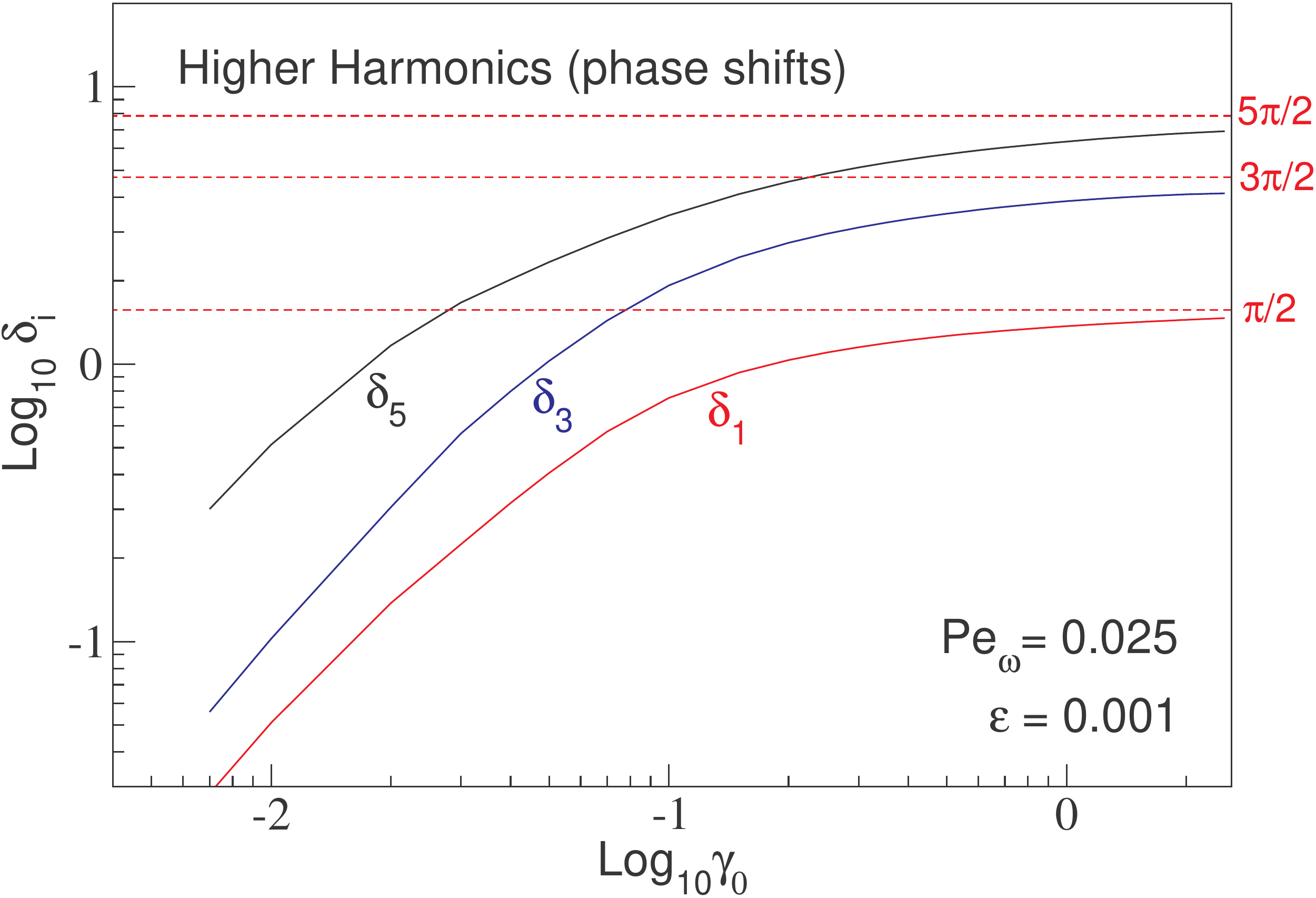}
\caption{{\color{red}(Theory)} 
The theoretical values for the first, third and fifth phase shift  
which contribute to the nonlinear stress response shown in Figure.\ref{th_f2}. 
It should be noted that the phases are only physically meaningful, i.e. contribute 
significantly to the total $\sigma(t)$ signal, when the corresponding intensities 
shown in Figure.\ref{th_f6} are non-negligable. 
($\epsilon=10^{-3}$ and $Pe_{\omega}=0.025$). 
}
\label{th_f7}
\end{figure}

For $\gamma_0>0.1$ the dissipation $G''_1$ becomes larger than $G'_1$, indicating a crossover from predominately elastic to predominately viscous response, and both functions exhibit a monotonic 
decay. For values of the strain amplitude larger than unity a regime of asymptotic decay is entered,   
characterized by a well defined power law dependence. 
We find that the numerically generated data are well fitted by power laws $G''_1\sim \gamma_0^{-\nu}$ and $G'_1\sim\gamma_0^{-\nu'}$, with $\nu=0.65(\pm 0.02)$ and $\nu'=2\nu$. 
Moreover, calculations performed at various frequencies show that the exponent values are independent 
\begin{figure}[!t]
\hspace*{0.3cm}\includegraphics[width=8.3cm]{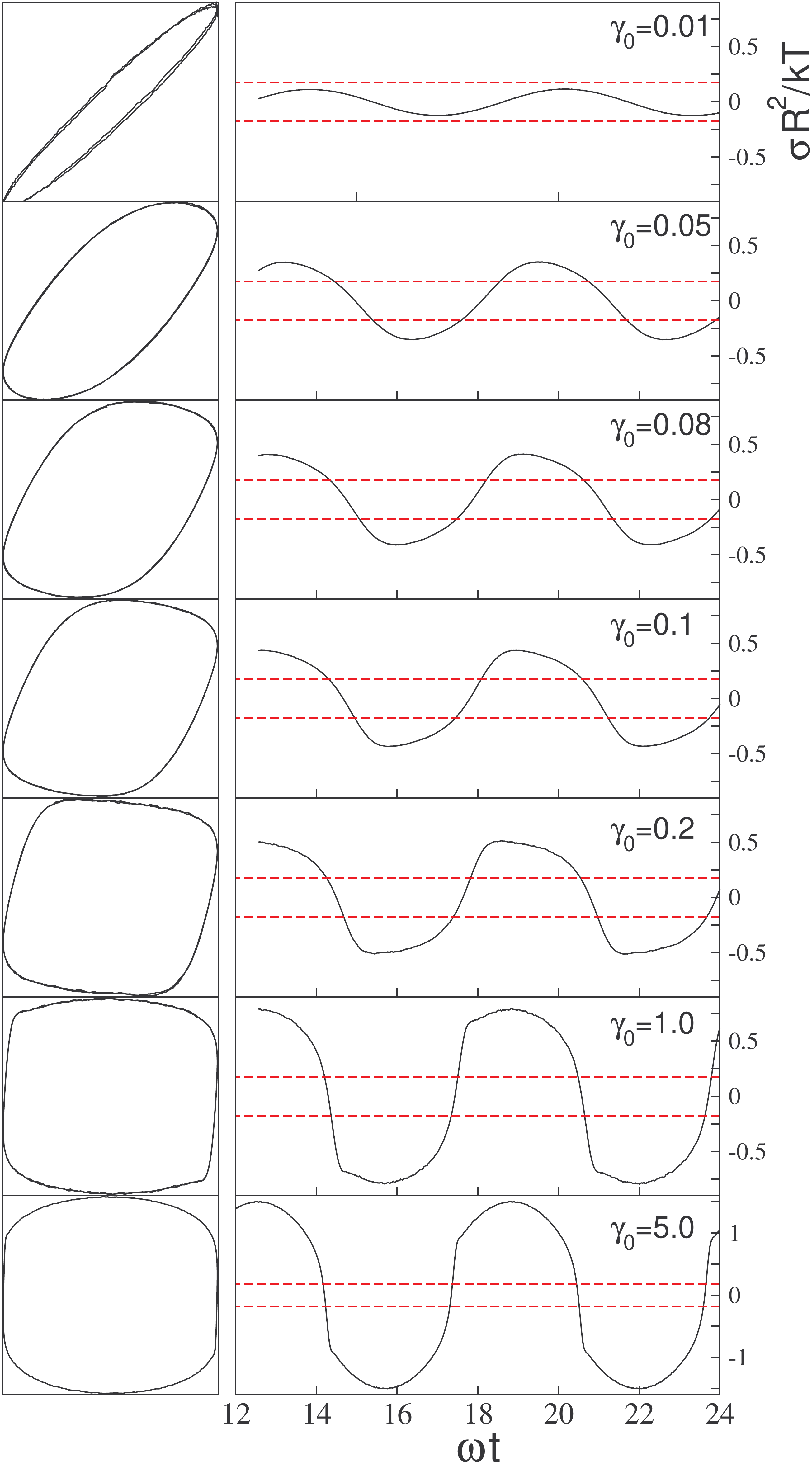}
\caption{{\color{red}(Simulation)} 
The stress response measured in Brownian dynamics simulations of a binary hard-disc mixture 
under oscillatory flow. A size ratio $1:1.4$ was used to supress crystallization. 
The considered strain amplitudes range from $\gamma_0=0.01$ to $\gamma_0=10$. 
The left column of figures shows the associated Lissajous 
curves illustrating the nonlinear character of the response. 
The simulations are performed at $\phi_{\rm tot}=0.81$ (slightly beyond  
the glass transition, according to our simulation estimates). 
The Peclet number is $Pe_{\omega}=0.05$). 
}
\label{sim_f1}
\end{figure}

of $\omega$ and thus seem to represent a universal aspect of the asymptotic decay within the schematic model.  
While the numerical findings are suggestive of a universal exponent $\nu$, analytical calulation of 
its precise value has so far proved elusive. 
The primary difficulty in extracting $\nu$ from the theory is that, even in the asymptotic regime, 
the correlator $\Phi(t,t')$ retains a residual dependence upon the waiting time $t'$ which does 
not yield readily to analytic treatment. 
The integral for the stress (\ref{non-tti}) thus consists of a complicated superposition of 
correlators for different values of $t'$. 

An important numerical prediction of the schematic model is that the exponents dictating the 
decay of $G'$ and $G''$ are related, to within numerical accuracy, by 
a factor of $2$. The relation $\nu'=2\nu$ has been observed in experimental studies of a variety of 
soft materials (see subsection \ref{exp_results} for more details on this point) and it would therefore be of considerable interest to investigate this apparent prediction of our model in more depth. 
The relationship $\nu'=2\nu$ is found also in simple nonlinear Maxwell models \cite{faraday,miyazaki3},  and is thus not particularly surprising. Such models inevitably predict the trivial exponents $\nu=1$ and $\nu'=2$. 
The numerical data obtained by Miyazaki {\em et al}. \cite{miyazaki3} from solving their microscopic 
MCT theory is consistent with the exponent relation $\nu'=2\nu$, but predicts a value of $\nu=0.9$ (obtained by fitting the numerical data), which differs somewhat from the experimental value 
$\nu=0.7$ for PMMA colloids presented in the same work. 
Whether the value $\nu=0.9$ from the MCT calculations of \cite{miyazaki3} is influenced by the additional isotropic approximations employed remains unclear.   

The coefficients $G'_1$ and $G''_1$ discussed above describe the response at the fundamental frequency. 
We now consider the contribution of the higher harmonic terms to the stress signal. 
Due to the $\dot\gamma(t)=-\dot\gamma(t)$ symmetry of $\sigma(t)$, even coefficients in the Fourier series (\ref{fourier_series_phase}) are identically zero within the schematic theory (a condition 
which provides a useful check for our numerical algorithms). 
In Fig.\ref{th_f6} we show the intensities of the odd harmonics (normalized by $I_1$) up to 
$n=9$ for a glassy state, obtained by applying a discrete Fourier transform to the time series $\sigma(t)$. 
For very small amplitudes $\gamma_0<0.01$ the numerical solution of the equation of motion for 
$\Phi(t,t')$ becomes unreliable, as the structural relaxation time exceeds the range of the numerical 
\begin{figure}
\hspace*{0.3cm}\includegraphics[width=8.cm]{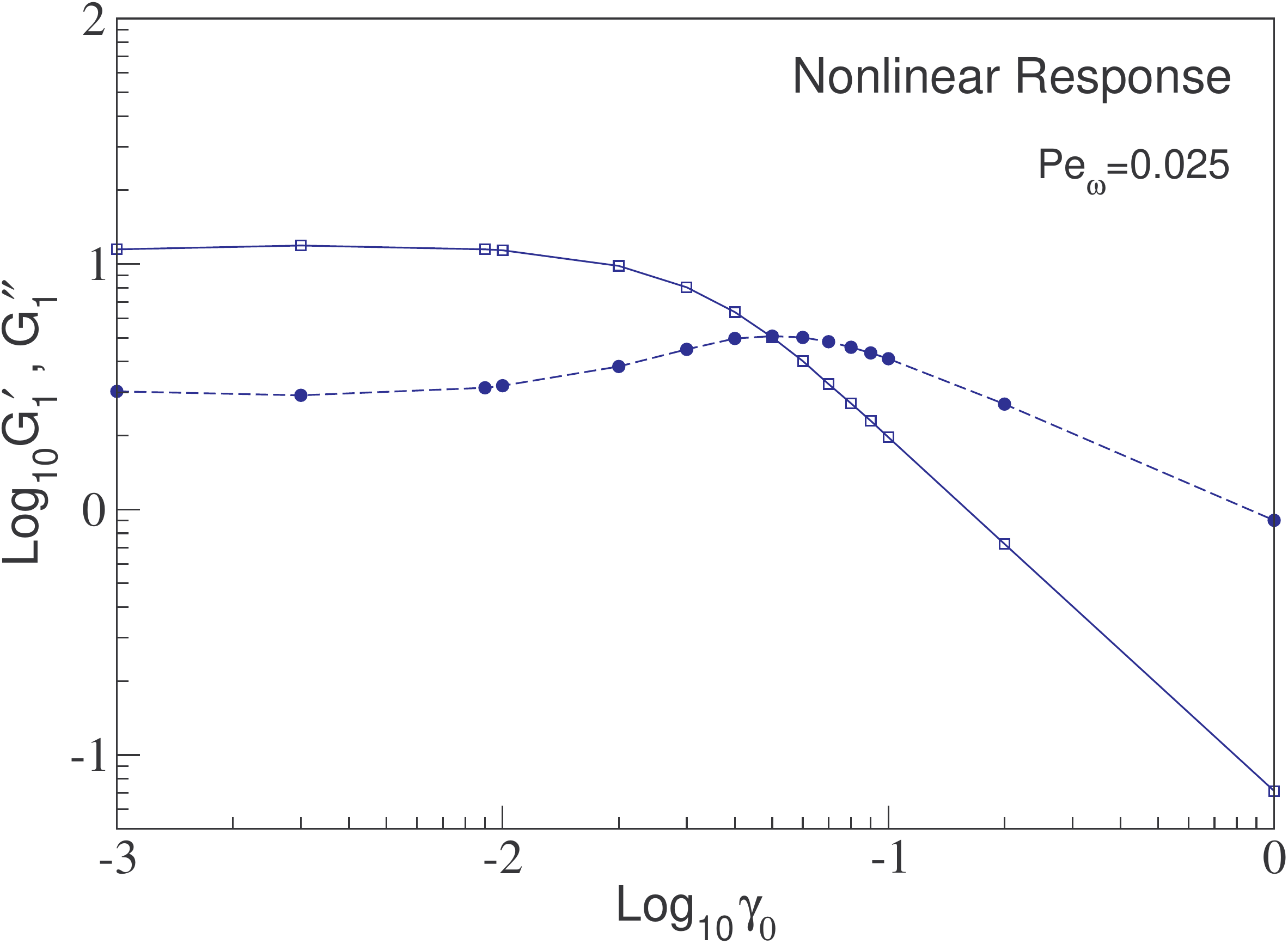}
\caption{{\color{red}(Simulation)} 
The simulation $G_1'$ and $G_1''$ as a function of strain for two dimensional 
volume fraction $0.81$ and Peclet number $Pe_{\omega}=0.05$. 
For large values of $\gamma_0$ the moduli obey the power laws 
$G''_1\sim\gamma_0^{-0.68}$ and $G'_1\sim\gamma_0^{-1.45}$.
}
\label{sim_f2}
\end{figure}

grid upon which the oscillating function $\Phi(t,t')$ can be resolved. 
Data are thus presented for $\gamma_0>0.01$ where accurate converged solutions can be obtained. 
At strain amplitude $\gamma_0=0.01$, only $I_3$ contributes significantly to the signal (around $3$\%). 
As the $\gamma_0$ is increased beyond $0.01$, the increasing influence of  $I_3$ is accompanied by 
the appearance of terms $I_5$, $I_7$ (beyond $\gamma_0=0.03$) and $I_9$ (beyond $\gamma_0=0.07$). 
Although intuitive, it is not clear {\em a priori} that the higher harmonics must neccessarily appear 
in sequence $n=3,5,7,\cdots$ upon increasing the amplitude. All of the $I_{n>3}$ exhibit a maximum in the range $0.3<\gamma_0<1$ and by $\gamma_0=1$ contribute approximately half of the total signal, 
$\sum_{n>1}I_n/I_1\sim 0.5$. 
The maximum and subsequent decay of the higher harmonics has also been observed in \cite{kallus}.

Complementary to the higher harmonic intensities are the phase shifts $\delta_n$ shown in Fig.\ref{th_f7}. It should be noted that  the phases are only physically meaningful for amplitudes 
at which the corresponding intensity is significant. 
As $\gamma_0$ is increased towards unity the phases saturate to the asymptotic values 
$\delta_1=\pi/2, \delta_3=3\pi/2$ and $\delta_5=5\pi/2$. 
The higher harmonic contributions $I_n$ and $\delta_n$, which contribute for $\gamma_0>0.01$, describe 
the distortion of $\sigma(t)$ close to the yield stress (c.f. Figs.\ref{th_f3} and \ref{th_f4}).


\subsection{Simulation results}
In Fig.\ref{sim_f1} we show the stress response measured in our Brownian dynamics simulations of a binary hard-disc mixture.
As the strain amplitude is increased the simulated stress evolves from a linear to a nonlinear response for 
$\gamma_0>0.03$. 
Consistent with the theoretical results shown in Fig.\ref{th_f3}, the time dependent signal becomes 
distorted away from a pure sinusoid when the peak of $\sigma(t)$ encounters the dynamic yield stress.  
Although the Peclet number $Pe_{\omega}=0.05$ is close to that employed in the theoretical calculations used to generate Fig.\ref{th_f3}, the onset of the yield stress clipping effect, already manifest in Fig.\ref{th_f3}, is not clear in the form of $\sigma(t)$ shown in Fig.\ref{sim_f1}. 
It would therefore seem likely that considerably smaller values of the Peclet number are required to observe this effect in our two-dimensional simulations. 
Nevertheless, the general form of the nonlinear stress is very similar, on a qualitative level, to that predicted by the schematic model in Fig.\ref{th_f3}. 
Both simulation and theory exhibit a flattened and asymmetric peak which is skewed to the left. 

\begin{figure}[!t]
\hspace*{0.3cm}\includegraphics[width=8.5cm]{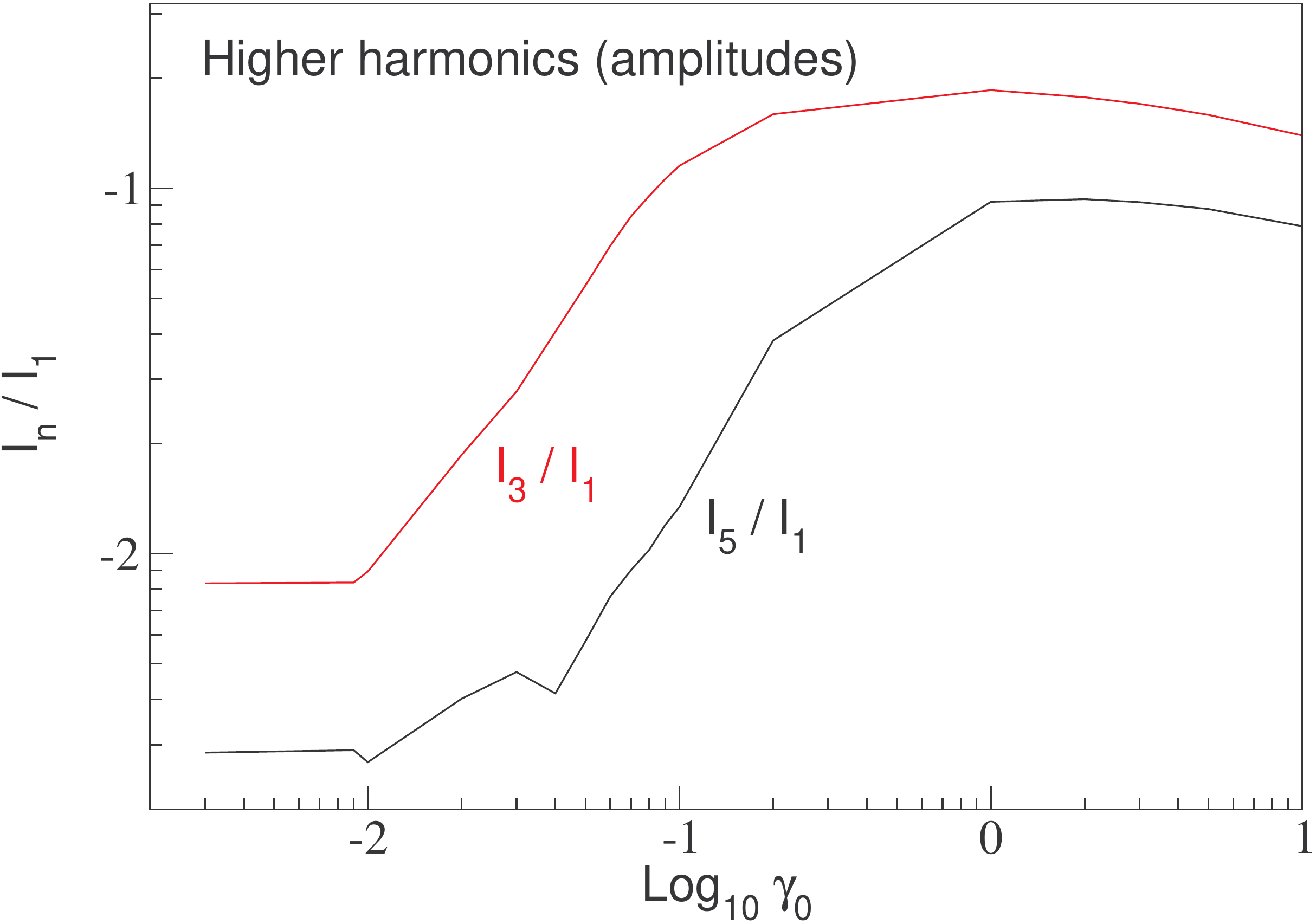}
\caption{{\color{red}(Simulation)} 
The normalized intensities of the third and fifth harmonic contributing to the nonlinear 
stress response shown in Figure.\ref{sim_f1}.  
}
\label{sim_f3}
\end{figure}

\begin{figure}[!t]
\hspace*{0.3cm}\includegraphics[width=8.3cm]{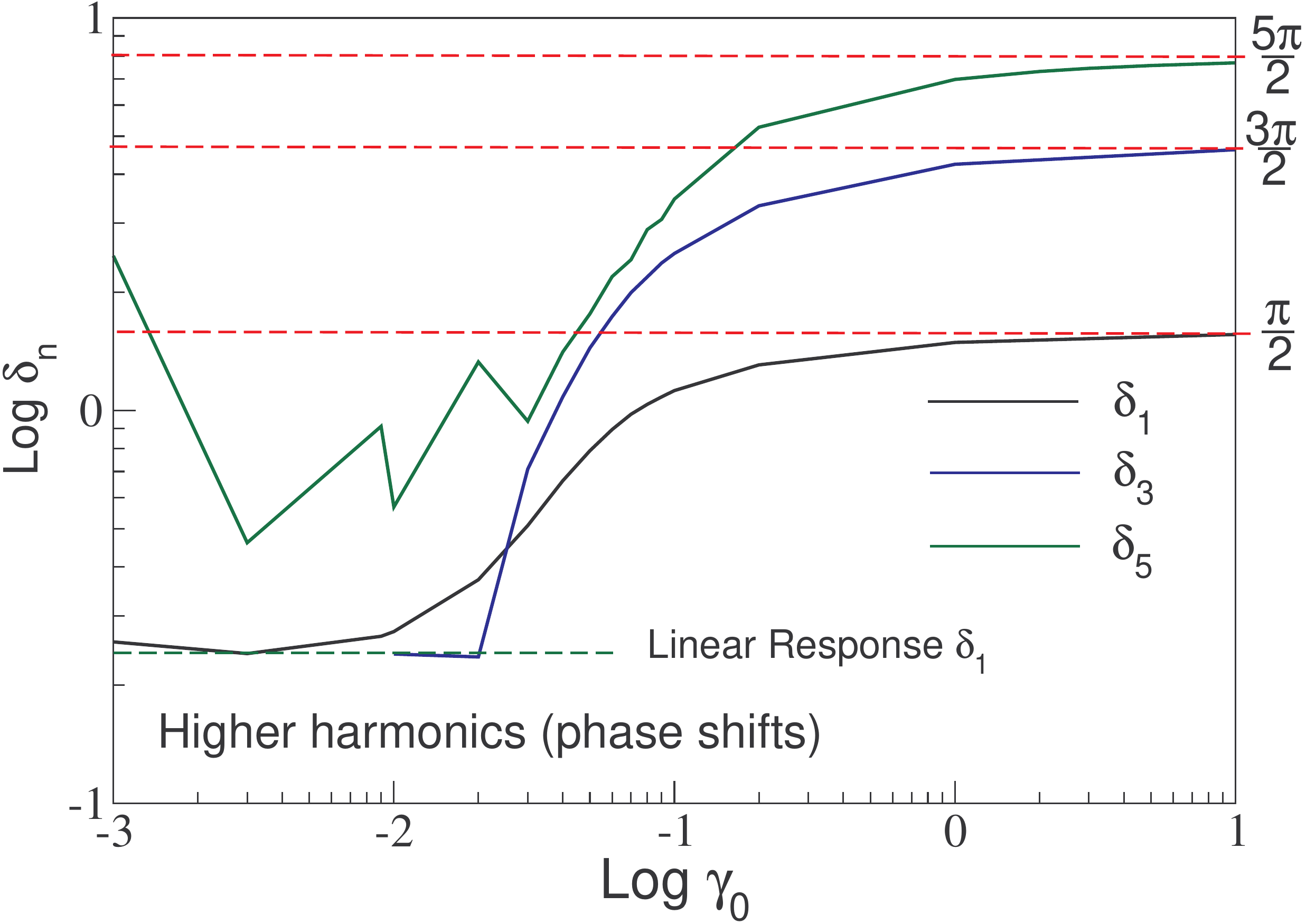}
\caption{{\color{red}(Simulation)} 
The phase shifts of the first and third harmonic contributing to the nonlinear 
stress response shown in Figure.\ref{sim_f1}. 
}
\label{sim_f4}
\end{figure}

In order to analyze more closely the stress signal we show in Figs.\ref{sim_f2}, \ref{sim_f3} and 
\ref{sim_f4} the fundamental coefficients, higher harmonic intensities and phase shifts, respectively. 
The dependence of $G'_n$ and $G''_n$ on $\gamma_0$ is strongly reminiscent of that predicted by the schematic model (Fig.\ref{th_f5}). 
Within the range $0.01<\gamma_0<0.06$ the system begins to deform plastically leading to a reduction 
in $G'_1$ and an increase in $G''_1$ (and cross at $\gamma_0=0.05$), reflecting the increasing importance of dissipative processes. 
The height of the peak in $G''_1$ is rather less pronounced than that predicted by the schematic 
model.
Beyond $\gamma_0=0.06$ both $G'_1$ and $G''_1$ decrease monotonically, exhibiting an asymptotic power law decay which is well described by the power laws $G''_1\sim\gamma_0^{-0.68}$ and $G'_1\sim\gamma_0^{-1.45}$. Although these values do not satisfy perfectly the empirical relation 
$\nu'=2\nu$, the deviation of the simulation exponent ratio $\nu'/\nu=2.13$ may well be attributable 
to numerical error. Despite this discrepancy, the absolute value of the exponents $\nu=0.68$ and 
$\nu'=1.45$ compare well with those emerging from the schematic model ($\nu=0.65$ and $\nu'=1.3$). 
The schematic model considered in the present work would thus seem to be more realistic than either 
a simple Maxwell model ($\nu=1$ and $\nu'=2$) or the microscopic MCT approach of Miyazaki 
{\em et al}. ($\nu=0.9$ and $\nu'=1.8$), at least on the basis of our simulation results. 

In Fig.\ref{sim_f3} we show the intensities of the third and fifth harmonic as a function of 
$\gamma_0$. Higher order terms were found to be highly susceptable to the effects of statistical 
noise in the simulation data and have thus been omitted. 
Upon increasing the strain amplitude beyond $\gamma_0=0.03$ the system leaves the linear 
response regime and the contribution of the third harmonic grows. 
For strains exceeding around $0.1$ the fifth harmonic also begins to play a significant role in determining the stress response. 
In keeping with the schematic models predictions, both $I_3/I_1$ and $I_5/I_1$ exhibit a maximum, 
albeit more sharply peaked and shifted to slightly larger strain values approaching unity.  
The corresponding phase shifts also share the general features of the schematic model predictions, 
in particular, for large values of $\gamma_0$ we find that $\delta_1$,$\delta_3$ and $\delta_5$ 
saturate to $\pi/2$, $3\pi/2$ and $5\pi/2$, respectively. 
These results suggest that the series (\ref{fourier_series_phase}) reduces to 
\begin{eqnarray}
\sigma(t)= \gamma_0\sum_{n=1}^{\infty} I_{2n+1}(\omega)(-1)^n\cos((2n+1) \,\omega t),  
\label{fourier_series_limit}
\end{eqnarray} 
for large values of $\gamma_0$.    

\begin{figure}[t!]
\includegraphics[width=8.3cm]{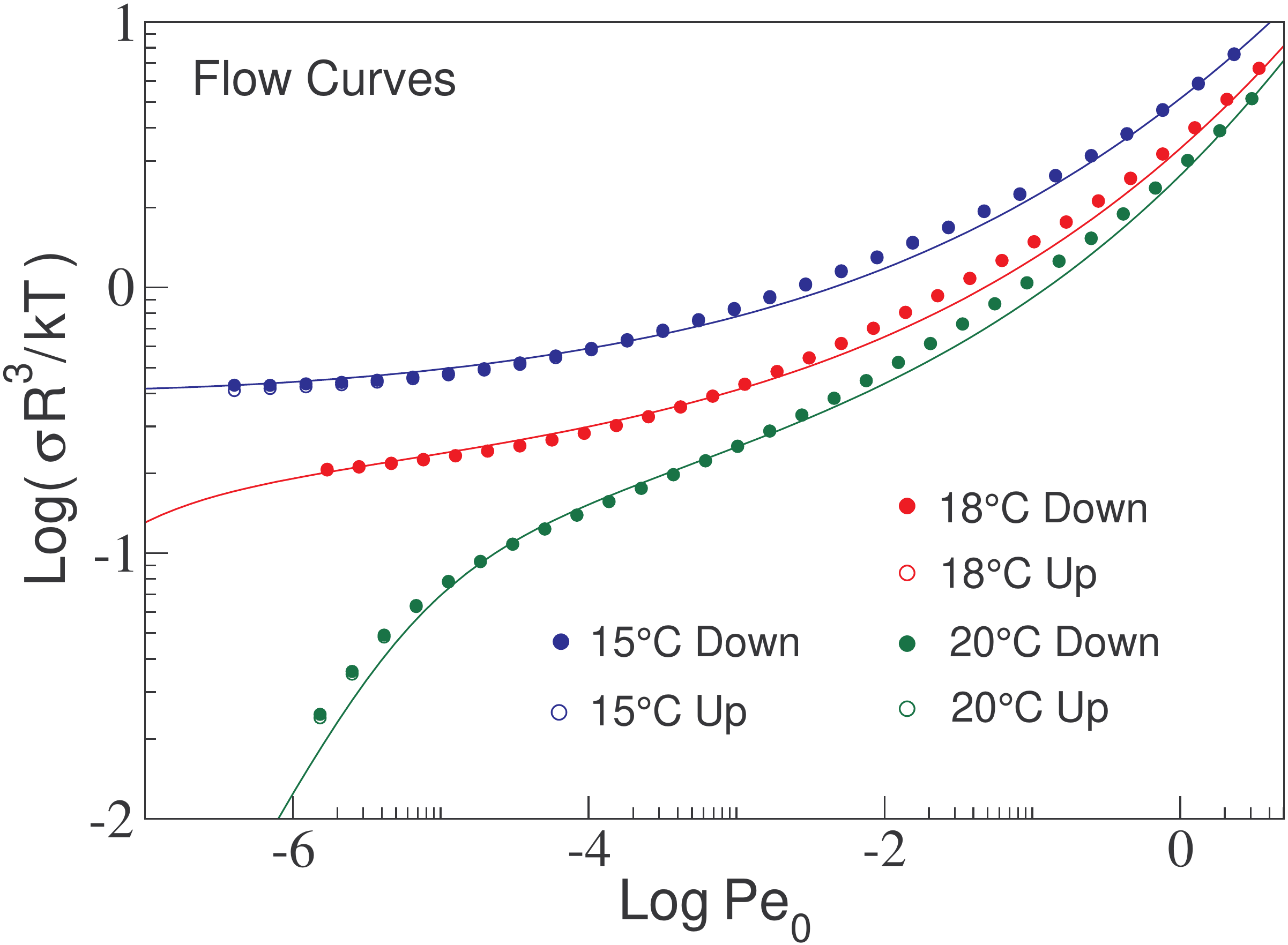}
\caption{{\color{red}(Experiment)} 
Symbols: The experimentally measured flowcurves for three different temperatures, 
$T=20$ ($\phi_{\rm eff}=0.57$), $T=18.4$ ($\phi_{\rm eff}=0.60$) and $T=15.1$ 
($\phi_{\rm eff}=0.65$). 
Lines: Theoretical fits to the data using the parameters in table \ref{parameters}
}
\label{exp_f5}
\end{figure}

\subsection{Experimental results}\label{exp_results}

\subsubsection{Flow curves and linear response moduli}
\begin{figure}[!t]
\includegraphics[width=8.3cm]{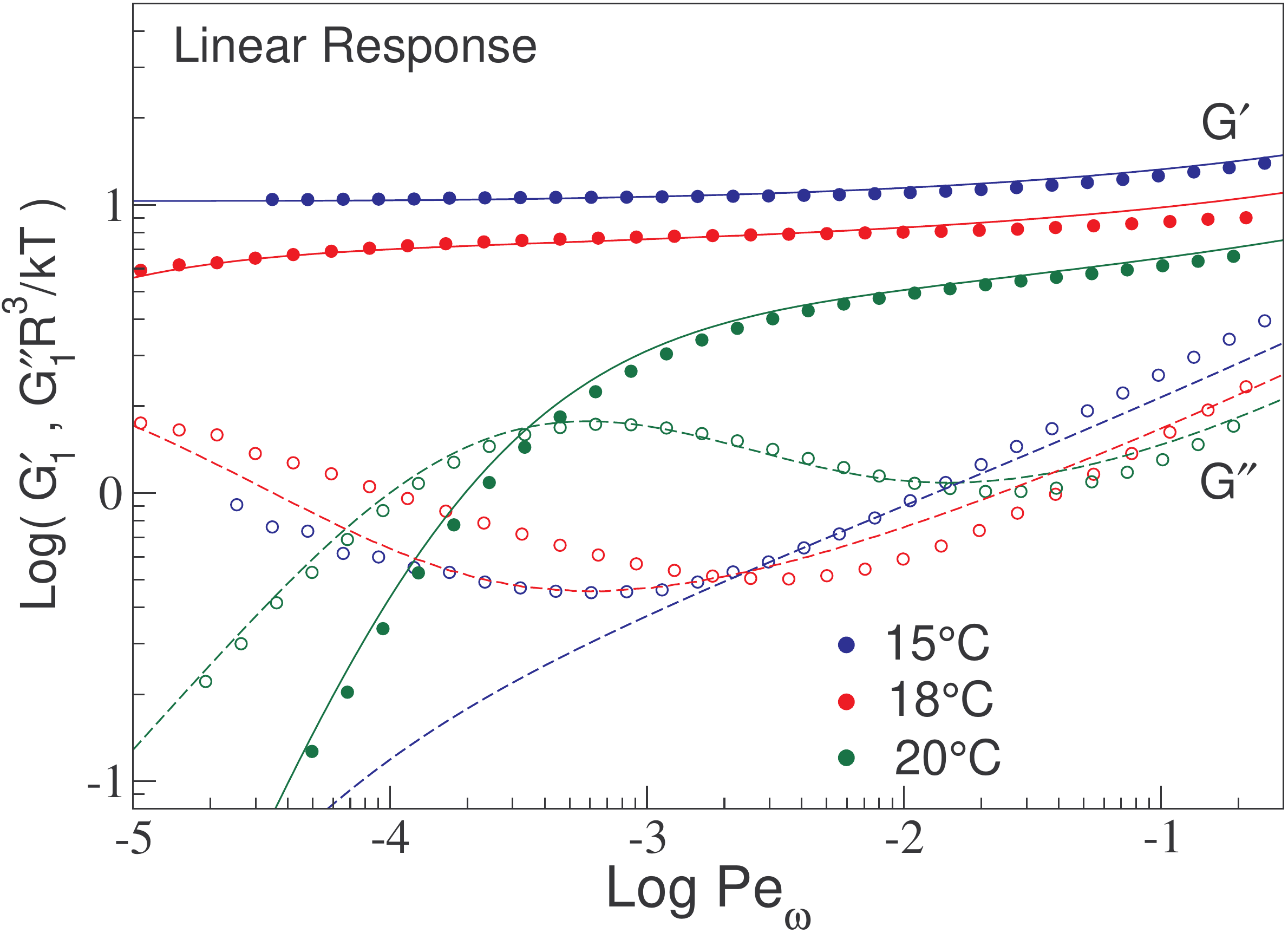}
\caption{{\color{red}(Experiment)} 
Symbols: The experimentally measured linear response moduli for three different 
temperatures, $T=20$ ($\phi_{\rm eff}=0.57$), $T=18.0$ ($\phi_{\rm eff}=0.60$) and $T=15.0$ 
($\phi_{\rm eff}=0.65$).
Lines: Theoretical fits to the data using the parameters in table \ref{parameters}
}
\label{exp_f4}
\end{figure}

In Fig. \ref{exp_f5} we show the experimentally measured flow curves and in Fig. \ref{exp_f4} the linear response moduli for three different temperatures (corresponding to three different volume fractions). 
Reduced units are employed for both the control parameters
\begin{eqnarray}
\mathit{Pe_0}=\frac{\dot{\gamma}R_H^2}{\mathit{D_0}}, 
\hspace*{1cm}
\mathit{Pe_{\omega}}=\frac{\omega R_H^2}{\mathit{D_0}},  
\notag
\end{eqnarray}
(where $D_0$ is obtained from the solvent viscosity using Stokes' law)
as well as for the shear stress and the moduli 
\begin{eqnarray}
\sigma_{\rm red} = \frac{\sigma R_H^3}{k_BT},
\hspace*{0.5cm}
G^{\prime}_{\rm red} = \frac{G^{\prime}R_H^3}{k_BT},
\hspace*{0.5cm}
G^{\prime\prime}_{\rm red} = \frac{G^{\prime\prime}R_H^3}{k_BT}.
\notag
\end{eqnarray}
For small Peclet numbers the flow curve measured at $20^{\circ}\,C$ (corresponding to $\phi_{\rm eff}=0.57$) shows a first Newtonian plateau. 
As $Pe_0$ is increased we observe a decrease in viscosity, followed by a second Newtonian plateau, both 
of which are typical for a shear thinning fluid. 
The viscoelastic character of the sample at $20^{\circ}\,C$ is demonstrated by the linear response moduli shown in Fig.\ref{exp_f4}. For intermediate frequencies $G^{\prime}$ and $G^{\prime\prime}$ cross, 
indicating the presence of a structural $\alpha$-relaxation process. 
 
The flow curve measured at the lower temperature $18^{\circ}\,C$ ($\phi_{\rm eff}=0.60$) displays a more pronounced plateau region. However, for the lowest frequencies investigated a slight decrease from 
the plateau is evident, suggesting the existence of an $\alpha$ relaxation time which has shifted out of the experimental frequency window.  
The corresponding linear moduli shown in Fig.\ref{exp_f4} show a distinctive plateau region of $G^{\prime}$ at intermediate values of $\mathit{Pe_{\omega}}$, followed by a decrease for  small $\mathit{Pe_{\omega}}$ values, consistent with the existence of a crossover point and, therefore, 
a fluid relaxation. 
This is expecially apparent in $G''(\omega)$ which continues to rise as the frequency is decreased. 
Extrapolation of the measured data to lower frequencies suggests $\mathit{Pe^{\rm cross}_{\omega}} \approx 10^{-6} - 10^{-7}$, where $G'$ and $G''$ cross. 
For the lowest temperature investigated, 15$^{\circ}\,C$ ($\phi_{\rm eff}=0.65$), the flow curve exhibits a constant plateau down to the lowest values of $Pe$ and the storage moduli remains constant at low 
$\mathit{Pe_{\omega}}$. 
The sample at 15$^{\circ}\,C$ may thus be considered as a glass for which additional `hopping' processes lead to an 
increase in $G''$ at low frequencies. 


\begin{figure}[t]
\includegraphics[width=7cm]{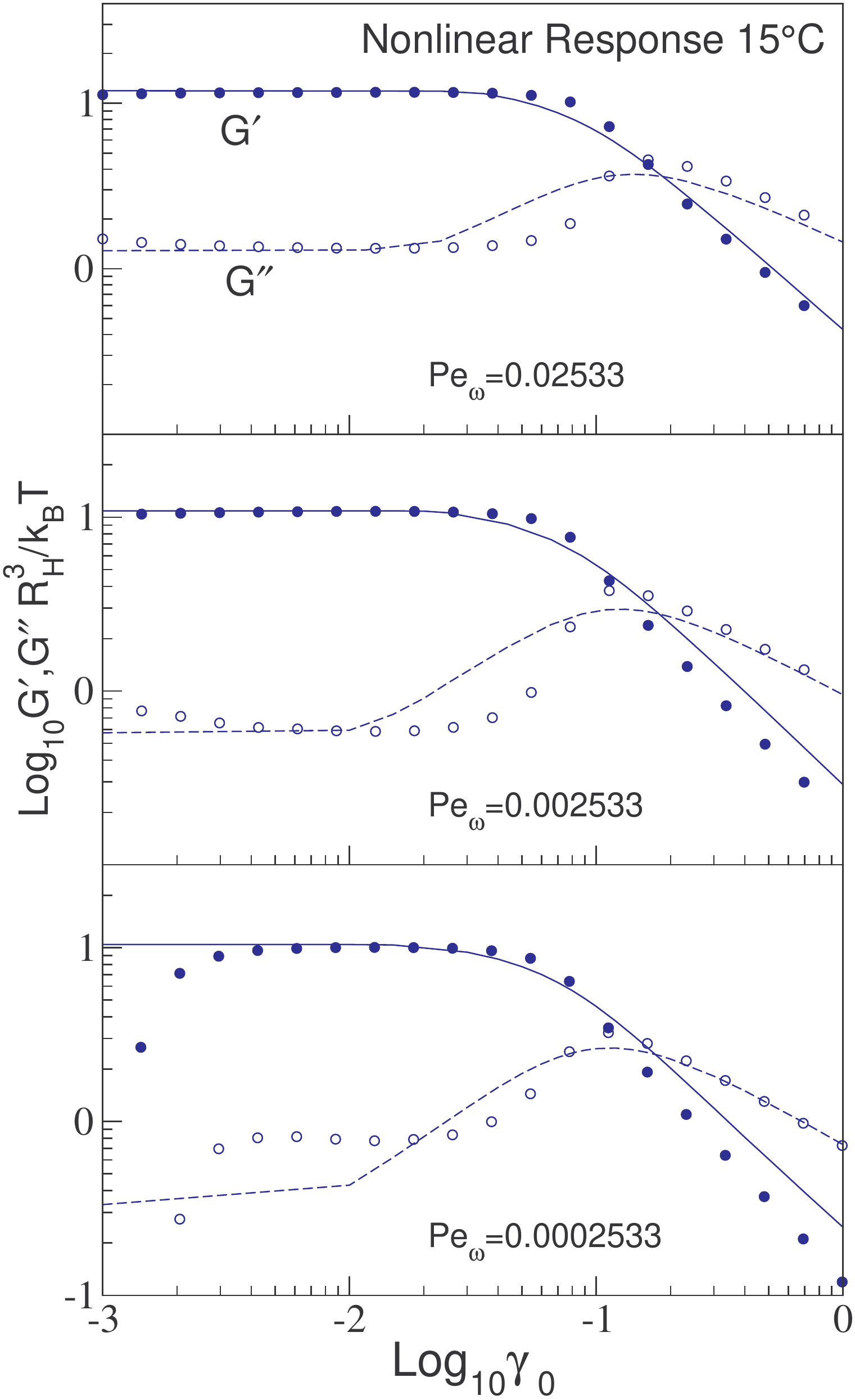}
\caption{{\color{red}(Experiment)} 
Strain sweeps at three different frequencies for the temperature $15.0^\circ C$. 
Circles: Experimental data. Lines: Theoretical fits using the parameter 
set $(\epsilon,v_{\sigma},\Gamma,\gamma_c,\tilde x)$ obtained by simultaneously 
fitting the flow curves (Fig.\ref{exp_f5}) and linear response moduli 
(Fig.\ref{exp_f4}).
}
\label{exp_f6}
\end{figure}

The procedure by which experimental data may be fit using the schematic F$_{12}^{\dot{\gamma}}$ model of \cite{faraday} is already well documented \cite{winter}. 
Fitting the experimental data for the flow curve and linear moduli using the present schematic model proceeds analogously. 
For a given volume fraction, a fixed set of model parameters may be found which fit both the flow curve and the linear moduli. 
It is thus possible to determine the separation parameter $\epsilon$, the vertex $v_{\sigma}$ and the decay rate $\varGamma$. The parameter $\gamma_c$ is obtained as an additional parameter for the description of the flow curve. The high frequency viscosity $\eta_{\infty}^{\omega}$ is only important for the frequency spectrum (and is connected with the high shear viscosity $\eta_{\infty}^{\dot{\gamma}}$ via $\eta_{\infty}^{\dot{\gamma}} = \eta_{\infty}^{\omega} + v_{\sigma}/(2\varGamma)$ \cite{crassous}). Fixing the parameters by these two experiments in the linear viscoelastic and the steady state determines all information needed to calculate the nonlinear oscillatory behaviour (parameters are summarized in table \ref{parameters}). Therefore, the experimental data sets of the deformation sweeps (Figs. \ref{exp_f6} - \ref{exp_f8}) and the oscillatory time tests (Figs. \ref{exp_f2} and \ref{exp_f3}) are solely described by the theory of Section III and the fixed parameter sets obtained by fitting Figs. \ref{exp_f4} and \ref{exp_f5} without further modification. 
In this sense, the theoretical results to be presented for large amplitude oscillatory shear are predictions, as no further fitting is required.


\subsubsection{Nonlinear 
response}

\begin{figure}[t]
\includegraphics[width=7cm]{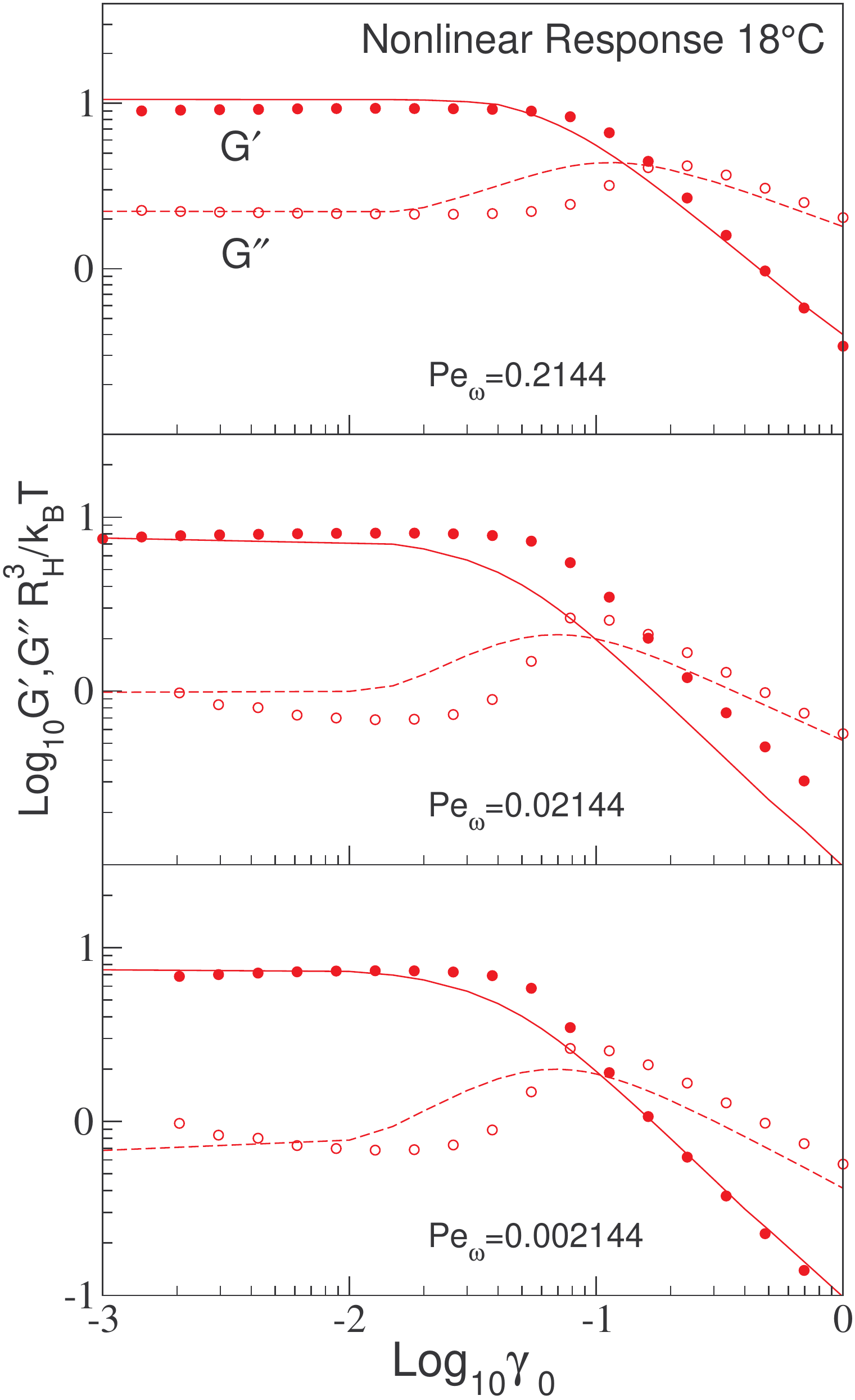}
\caption{{\color{red}(Experiment)} 
Strain sweeps at three different frequencies for the temperature $18.0^\circ C$. 
Circles: Experimental data. Lines: Theoretical predictions.  
}
\label{exp_f7}
\end{figure}

\begin{figure}[t!]
\includegraphics[width=7cm]{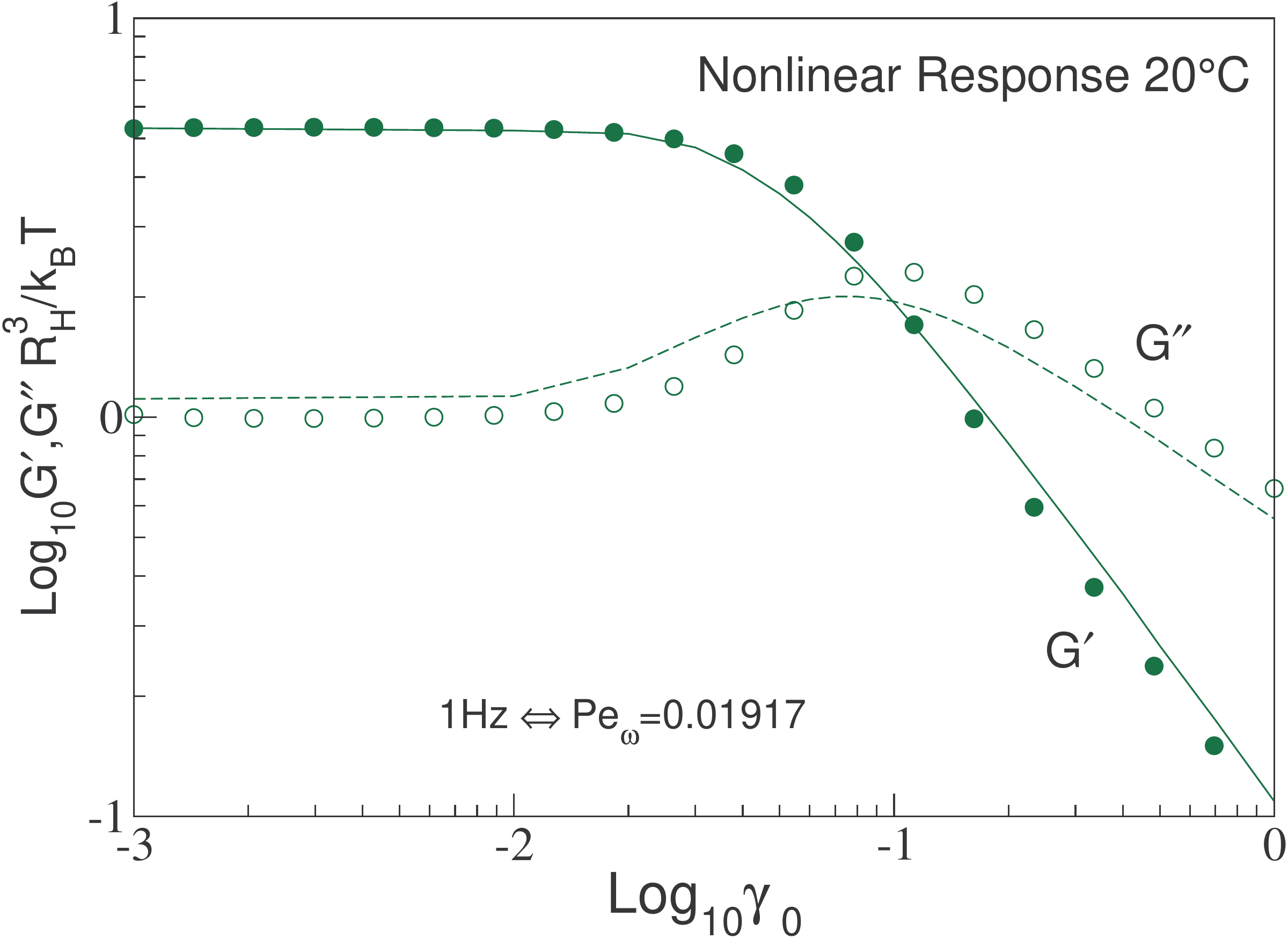}
\caption{{\color{red}(Experiment)} 
Strain sweeps at a fixed frequency $Pe_{\omega}=0.01917$ for the temperature 
$20^\circ  C$. 
Circles: Experimental data. Lines: Theoretical predictions. 
}
\label{exp_f8}
\end{figure}

\begin{figure}[t!]
\includegraphics[width=8.3cm]{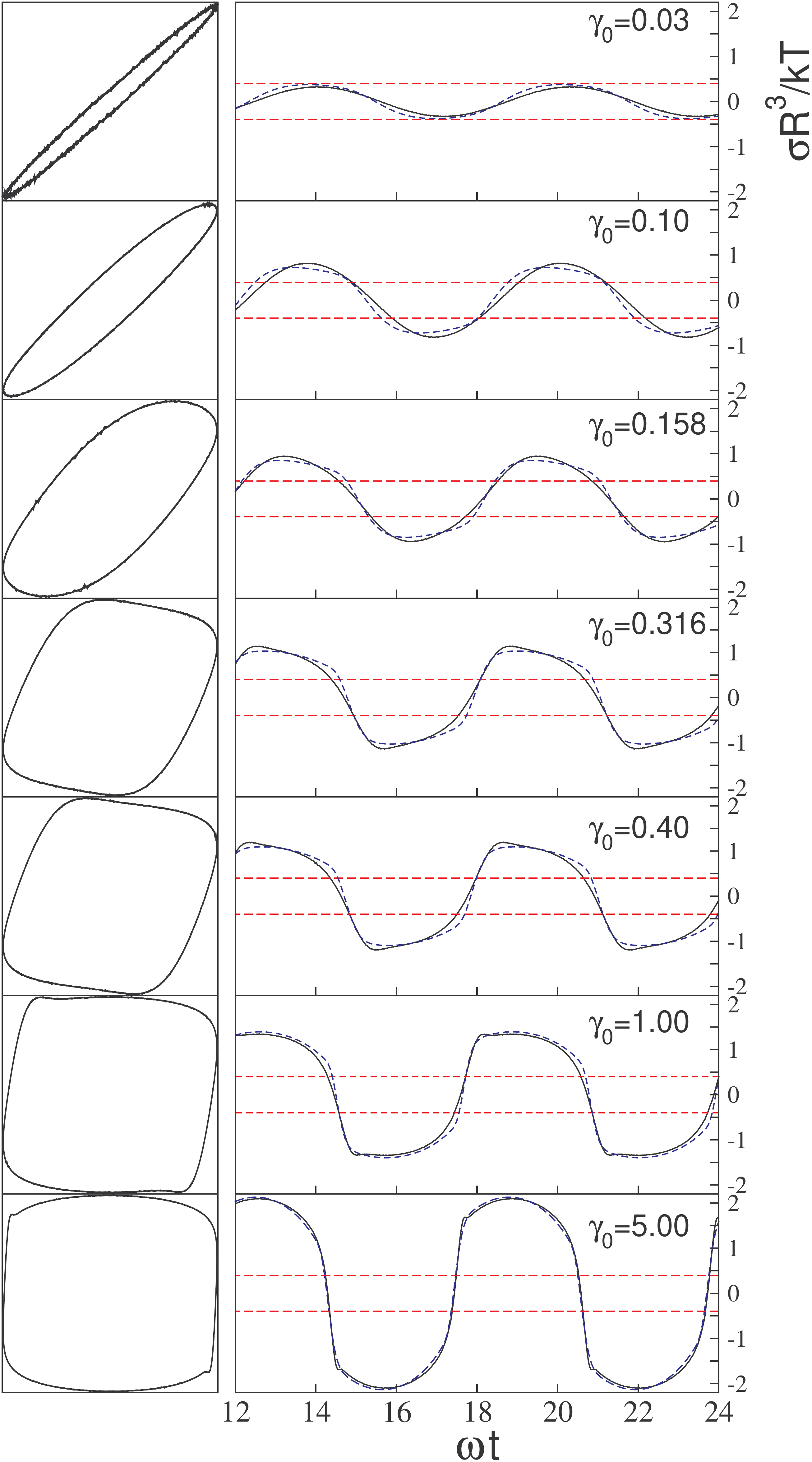}
\caption{{\color{red}(Experiment)} 
The stress response measured in LAOS experiments for strain amplitudes from $\gamma_0=0.03$ to 
$\gamma_0=5$ (black line) and the associated Lissajous figures illustrating the nonlinear character of the response. The experiments are performed at $T=15.1^\circ C$ a glassy state and at a frequency of $1Hz$ (corresponding to $\mathit{Pe_{\omega}}=0.02533$). At $\gamma_0=0.03$ the response is almost entirely elastic, emphasising the proximity of the quiescent state to the glass transition. At $\gamma_0=5$ the system is almost purely viscous. The increase in dissipation with increasing $\gamma_0$ is reflected in the increasing area enclosed by the closed Lissajous curves. 
The yield stress is indicated by the broken red lines. 
Theoretical results are given by broken blue lines. }
\label{exp_f2}
\end{figure}

\begin{figure}[t!]
\includegraphics[width=8.3cm]{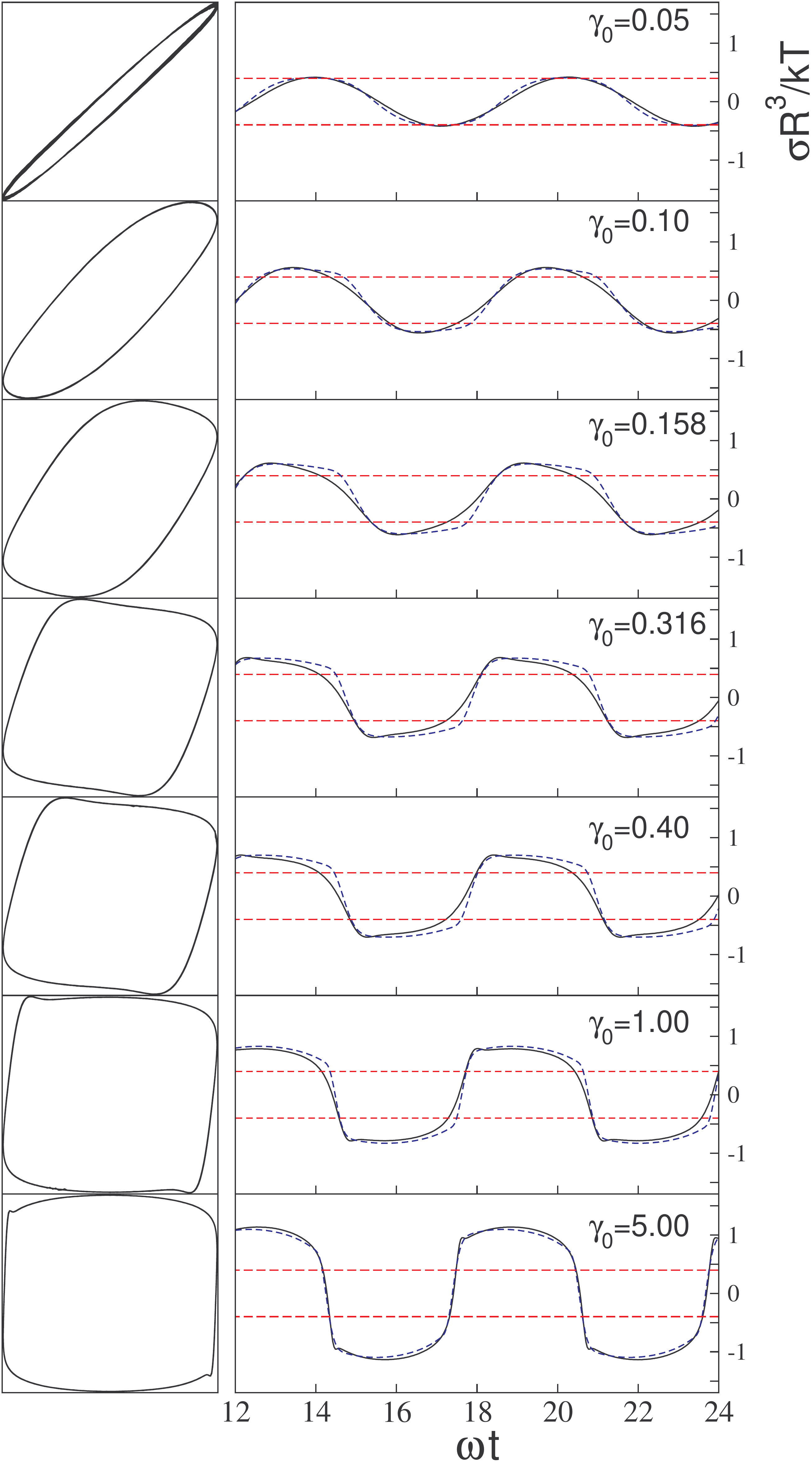}
\caption{{\color{red}(Experiment)} 
As in Fig.\ref{exp_f2} but at a frequency of $0.1Hz$ 
(corresponding to $Pe_{\omega}=0.002533$). Full black lines: Experimental data. 
Broken blue lines: Theoretical results. 
The yield stress is indicated by the broken red lines. 
}
\label{exp_f3}
\end{figure}

The nonlinear regime has been tested using two different experiments: deformation sweeps at 1, 0.1 and 0.01$\,$Hz (see Figs. \ref{exp_f6} - \ref{exp_f8}) and oscillatory time series measured at 1, 0.1 and 0.01$\,$Hz for various deformation amplitudes ranging from the linear to the non-linear regime (see Figs. \ref{exp_f2} and \ref{exp_f3}). The complete set of nonlinear oscillatory data is solely described by the schematic MCT employing the parameter sets determined by the fitting procedure described above. 

\begin{table}[!b]
 \begin{tabular}{|c|c|c|c|c|c|c|}
  
    \hline
    T [$^\circ$C] & $\phi_{\rm eff}$ & $\epsilon$ & $v_{\sigma}$ & $\Gamma$ & $\gamma_c$ & $\eta_{\infty}$ \\
   \hline
    20.0    & 0.57   & $-2.45\times 10^{-3}$ & 59 & 100 & 0.18  &42 \\
    18.0    & 0.60   & $\;\;-2.2\times 10^{-4}$ & 85 & 100 & 0.19  & 36\\
    15.0    & 0.65   & $\;\;\;\;\;5\times 10^{-5}$ & 115 & 105 & 0.28  & 24 \\

   \hline
    \end{tabular}
   \caption{Schematic model parameters obtained by fitting experimental data for the flow curves (Fig.\ref{exp_f5}) and 
   linear response moduli (Fig.\ref{exp_f4}). These parameters are then used to make theoretical prediction for large 
   strain amplitude oscillatory experiments.}
  \label{parameters}
\end{table}

In Fig.\ref{exp_f6} we show the results of strain sweep experiments at 15$^{\circ}\,C$ ($\phi_{\rm eff}=0.65$) for three different values of $\mathit{Pe_{\omega}}$. 
For strains up to around 1\% the system shows linear response behaviour, beyond which dissipation starts to increase, leading to a growth of $G^{\prime\prime}$ up to a maximum in the range 10-20\% strain. 
The growing dissipation and eventual crossing of $G^{\prime}$ and $G^{\prime\prime}$ as a function of $\gamma_0$ indicate the breaking of microscopic particle cages. 

For higher deformations $G^{\prime}$ and $G^{\prime\prime}$ display a power law decay.  
The exponents $\nu$ and $\nu'$ obtained at different temperatures and frequencies are given in table \ref{table2}

The theoretical predictions shown in Fig.\ref{exp_f6} are in excellent agreement with the experimental data and capture both the height and location of the maximum in $G^{\prime}$ rather well, although 
the departure from linear response appears to be less abrupt than in experiment, indicating a more gradual breaking of cages with increasing $\gamma_0$. 
We note that the discrepancy between theory and experiment in the linear response regime for the lowest frequency considered has its origins in the linear moduli fits presented in 
Fig.\ref{exp_f4}. For glassy states there occur additional physical relaxation processes at low frequency which are manifest in an upturn of the linear response $G^{\prime\prime}(\omega)$ at low frequencies and which are not captured by mode-coupling based theoretical approaches. 
Particularly significant is the excellent agreement between theory and experiment in the large strain regime for which the cage structure has been broken up by the applied flow. 
The power law decay of the experimental data is well described by the theoretical exponents $\nu=0.65$ and $\nu'=1.3$. 

Both the numerically obtained theoretical results and experimental data shown here demonstrate the exponent relation $\nu' \approx 2\nu$ (as in the theory of \cite{miyazaki3}). It is interesting to note that although this relation does not appear to be truly universal, broadly similar behaviour has been found for a variety of different materials, all of which show the deformation behaviour classified by Hyun \textit{et al.} \cite{Hyun02} as type III (weak strain overshoot). 
A few examples are e.g.~a Xanthan gum solution \cite{McKinley10} with $\nu' = -1.53$, $\nu=0.64$ and a ratio $\nu'/\nu=2.4$, anchor spreadable butter and promise spread which yield exponents 
$- 2.1<\nu'<-2.0$ with $\nu= -0.9$ and a ratio of $\nu'/\nu=2.3$, or the hard-sphere solution of Miyazaki \textit{et al.} \cite{miyazaki3} (PMMA spheres of 197nm in a mixture of decaline and cycloheptylebromide) which show $\nu = 0.7$ and $\nu'=1.4$. 

In Figs.\ref{exp_f7} and \ref{exp_f8} we show further strain sweep measurements for temperatures 
18$^{\circ}\,C$ ($\phi_{\rm eff}=0.60$) and 20$^{\circ}\,C$ ($\phi_{\rm eff}=0.57$). The overall level 
of agreement between theory and experiment is even better than that found for T=15$^{\circ}\,C$. In particular, the strain sweep measured at 20$^{\circ}\,C$ is very well described by the theory and, 
taken together with the results shown in Figs.\ref{exp_f5} and \ref{exp_f4}, demonstrate the accuracy 
with which the present schematic model may be used to describe the flow curve, linear moduli and strain sweep data of a dense colloidal fluid using a single fixed set of model parameters.  

For the oscillatory time series the schematic model calculations and the simulation are in good agreement with the experimental data. The direct comparison of theory and experiment for two frequencies (1$\,$Hz and 0.01$\,$Hz) respectively the Peclet numbers ($\mathit{Pe_{\omega}} = 0.025$ and 0.0025) is given in Figs. \ref{exp_f2} and \ref{exp_f3}. For small deformations, a linear viscoelastic behaviour is indicated by the nearly perfect sinusoidal output signal, but becomes distorted as the strain amplitude is increased. 
For $\mathit{Pe_{\omega}} = 0.025$ the stress signal displays flattened asymmetric peaks at intermediate values of $\gamma_0$, consistent with a regime of cage breaking around $\gamma_0=\gamma_c$. 
In contrast to the MCT predictions, the data show a pronounced dip at the top of the asymmetric peak. 
For high deformations, the peak shape approaches a semi-spherical shape with a vanishing but still visible dip or overshoot at the beginning edge of the peak. For the smaller frequency at $\mathit{Pe_{\omega}} = 0.0025$ indications of the effect of the merging $\sigma_{max}$ and the yield stress are found. The peaks show in the intermediate and high deformation range a more cut off shape, although the dip still remains, in contradiction to the MCT. The peak shape for middle $\gamma_0$ drops faster in the experiment as for the more box-like shape of the MCT time signals. However this shape is found in the experiment for high $\gamma_0$ as well. Alltogether the experiments and the theory fit well for all compared Peclet numbers and amplitudes, although some small deviations of the shape exist. The predictive character of the schematic MCT model for the oscillatory time series is remarkable, due to the fact that the shapes are not fitted but calculated with the parameter set defined from the flow curves and the frequency test in the linear viscoelastic regime.   

\begin{table}[!b]
 \begin{tabular}{|c|c|c|c|c|c|}
  
    \hline
    T [$^\circ$C] & $\phi_{\rm eff}$ & $\omega$ [Hz] & $\nu$ & $\nu'$ & $\nu'/\nu$  \\
   \hline
    15.0    & 0.65   & 1    & -0.633 & -1.262 & 1.99 \\ 
    15.0    & 0.65   & 0.1  & -0.742 & -1.404 & 1.89 \\     
    15.0    & 0.65   & 0.01 & -0.788 & -1.555 & 1.97 \\ 
    18.0    & 0.60   & 1    & -0.631 & -1.297 & 2.06\\
    18.0    & 0.60   & 0.1  & -0.712 & -1.424 & 2.00\\
    18.0    & 0.60   & 0.01 & -0.824 & -1.630 & 1.98\\             
    20.0    & 0.57   & 1    & -0.630 & -1.250 & 1.98\\
   \hline
    \end{tabular}
   \caption{Experimentally measured values of the exponents $\nu$ and $\nu'$ dictating the decay 
   of $G'$ and $G''$ as a function of $\gamma_0$ for large values of $\gamma_0$.}
  \label{table2}
\end{table}


\begin{figure}[!t]
\includegraphics[width=7.0cm]{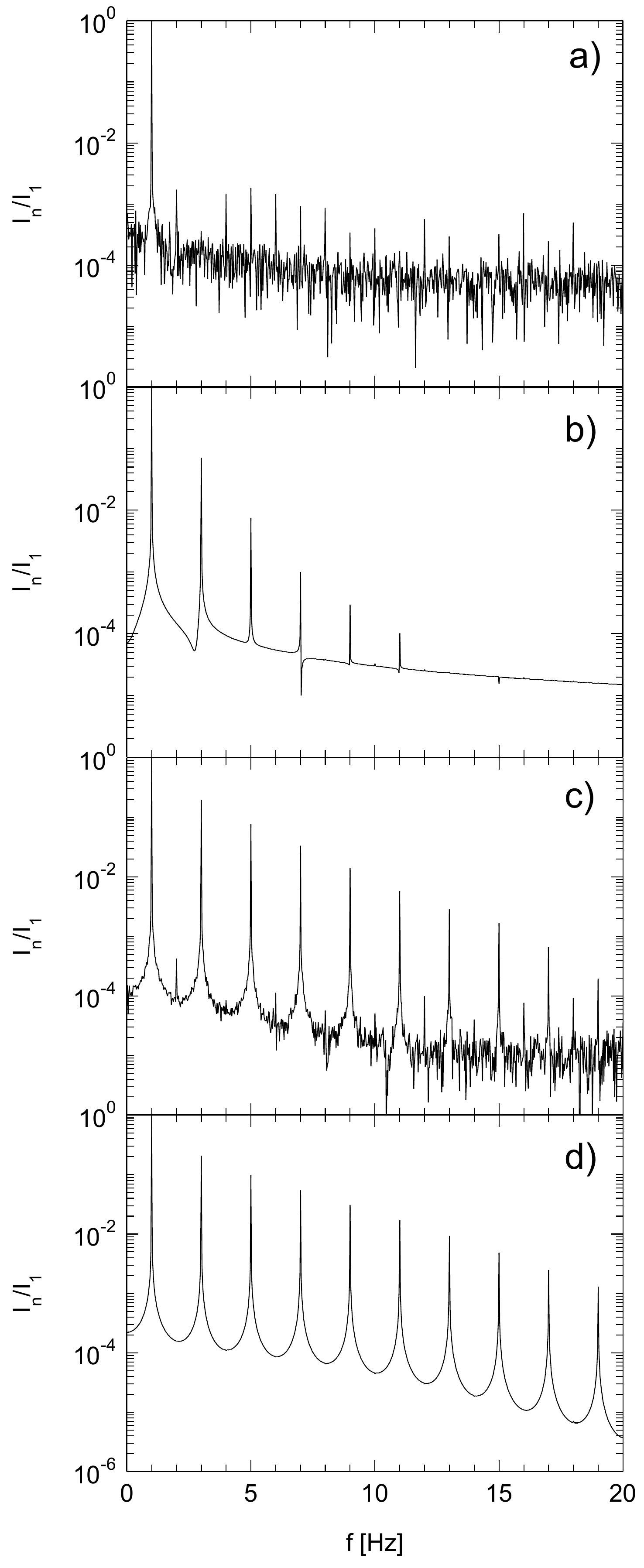}
\vspace{-0.25cm}
\caption{{\color{red}(Experiment and Theory)} 
FFT-spectra of the oscillatory time test signals. The experiment was performed at 15.1$^\circ\,$ and 1$\,$Hz.  $a)$ (experiment) and $b)$ (theory) are the FFT spectra at $\gamma_0 = 0.03$ and $c)$ (experiment) and $d)$ (theory) at an amplitude of $\gamma_0 = 1$.
}
\label{FFT}
\end{figure}

\subsubsection{Fourier analysis}
In order to provide a more detailed analysis of the time series shown in Figs.\ref{exp_f2} and \ref{exp_f3} we have calculated the intensities and phase shifts (see Eq.(\ref{fourier_series_phase})). 
The phase shifts are calculated from the real and imaginary parts of the Fourier transformed signal. 
Representative FFT-spectra are shown in Fig.~\ref{FFT} for the measurements made at at 15.1$^\circ\,$C, 1$\,$Hz and for amplitudes $\gamma_0 = 0.03$ and $1$. 
The spectrum demonstrates that at this low strain amplitude the higher harmonics do not contribute significantly and have an intensity only around 0.1\% of the basic harmonic $I_1$. 
Although peaks for even and odd harmonics are visible, their low intensities are not far away from the noise level. 
We therefore consider this measurement to be within the linear viscoelastic range. 

At the same conditions the MCT calculations show a rather different picture (see Fig.~\ref{FFT}b). Although the time signals for this $\gamma_0$ do not deviate strongly from the experimental equivalent (see Fig. \ref{FFT}), the effect in the Fourier spectrum is noticable. Up to the 11th harmonic all odd harmonics can be clearly separated from the base line, with $I_3/I_1$ around 1\%. 
Furthermore, in the theoretical FFT-spectrum an exponential decay of the intensities is apparent, a feature which first becomes evident in the exterimental data for $\gamma_0 \geq 0.158$. 
At the larger strain amplitude, $\gamma_0 = 1$, the MCT calculations and the experimental FFT-spectra are very similar (see Fig.~\ref{FFT}c and d). 
The present theory thus provides a qualitative description of the higher harmonics which becomes semi-quantitative 
as the strain amplitude becomes significant ($\gamma_0\sim 1$). 
The theory provides a sensible intperpolation between linear response and large amplitude regimes, both of which are 
captured accurately.

Some insight into the origin of the discrepancies at small excitation amplitude may be obtained 
by considering the theoretical predictions for the buildup of stress following the onset of steady 
shear flow (see also \cite{zausch}). 
Within the elastic regime (i.e. well below the 'stress overshoot' identifying the yield strain) the system should display Hookian behaviour with a clearly defined elastic constant. 
However, schematic model calculations of the stress for this protocol (either using the present model or the $F_{12}^{\dot\gamma}$ model of \cite{faraday}) exhibit devaiations from Hookian behaviour at small strains. 
This feature of the schematic models is related to the slowness of the $\beta$ decay of the transient density correlator onto the plateau.
We can thus speculate that the discrepancy between the Fourier spectra of theory and experiment at low values of $\gamma_0$ may be due to the excessively slow decay of the correlator to its plateau value, which is an inherent feature of any model based on the original schematic $F_{12}$ model 
\cite{MCTequations,goetze_zeit,sjoegren}.

\begin{figure}[!t]
\includegraphics[width=6.8cm]{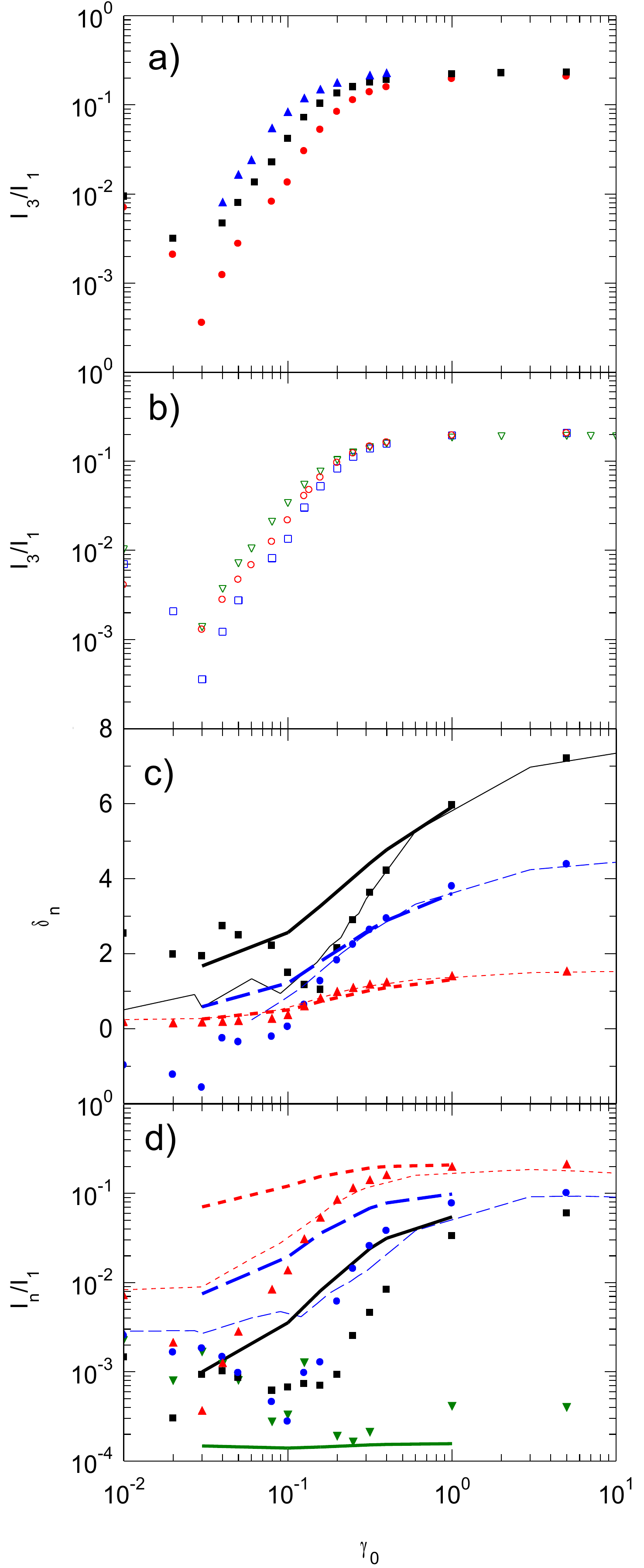}
\vspace{-0.25cm}
\caption{{\color{red}(Experiment and Theory)} $a)$ $I_3/I_1$ at $15.1^\circ\,$C for different frequencies (red circle: 1$\,$Hz, 
black square: 0.1$\,$Hz and blue triangles up: 0.01$\,$Hz). 
$b)$ Temperature dependence of $I_3/I_1$ at 1$\,$Hz: $15.1^\circ\,$C (blue squares), 18.4$^\circ\,$C (red circles) and 20.9$^\circ\,$C (green triangles down).
$c)$ Experimental phase shifts (symbols) for 15.1$^\circ$C at 1$\,$Hz. The simulation data is plotted versus $\gamma_0=3\gamma_0^{sim}$ as thin lines. The MCT calculations is given in thick lines. The phase shifts are $\delta_1$ (red triangles up and dotted lines), $\delta_3$ (blue circles and dashed lines) and $\delta_5$ (black squares and solid lines).
$d)$ Normalized intensities at 15.1$^\circ\,$C and 1$\,$Hz; symbols mark the experimental data, thick lines the FFT of the MCT time signals and thin lines the shifted simulation results: $I_2/I_1$ (green triangles down and solid line), $I_3/I_1$ (red triangles up and dotted lines), $I_5/I_1$ (blue circles and dashed lines) and $I_7/I_1$ (black squares and solid line). 
}
\label{exp_f9}
\end{figure}

In Fig.~\ref{exp_f9}a the amplitude dependence of the experimentally measured third harmonic is shown for three different frequencies in the glass (15.1$^\circ\,$C). It can be seen that as the frequency is reduced, the onset of the nonlinear regime, indicated by increase of the normalized third harmonic intensity, moves to lower values of $\gamma_0$. 
In Fig.~\ref{exp_f9}b we show the same quantity at a fixed frequency of 1Hz for three different temperatures. Surprisingly no strong influence of a change in the volume fraction is found, apart from  
a small deviation in the onset of the non-linear regime in the glassy state, which is shifted to 
higher deformations. The only significant deviation for the volume fractions considered occurs at intermediate deformations, as the starting and end values coincide.

The phase shifts are given for the measurement at 1$\,$Hz ($\mathit{Pe_{\omega}}= 0.025$) and 15$^{\circ}\,$C in Fig.~\ref{exp_f9}c. In the case of the fundamental $\delta_1$ the phase shift is found as expected to start at 0$^{\circ}$ for small amplitudes and to end at 90$^{\circ}$ for high deformations, which corresponds to a cosine. The curve progression of the experimental data and the MCT calculation fits perfectly for $\delta_1$. 
The theory deviates for the phase shifts of the higher harmonics due to the excessive contribution of non-linearities at lower $\gamma_0$. 
However, the limiting values for high deformations are found to coincide ($\frac{\pi}{2}$ for $\delta_1$, $\frac{3\pi}{2}$ for $\delta_3$ and $\frac{5\pi}{2}$ for $\delta_5$). 

We have also included the data from our two-dimensional simulations into Figs.~\ref{exp_f9}c and d. 
In order to obtain a reasonable comparison we found it neccessary to empirically multiply the strain employed in the simulations be a factor of three, $\gamma_0 = 3\cdot \gamma_0^{sim}$. 
That such an empirical rescaling is necessary is not surprising, given that only qualitative comparison is to be expected when comparing three dimensional experimental results with those of two dimensional simulations. 
It is therefore gratifying that the phases of the rescaled $\gamma_0^{sim}$ are found to describe the experimental data very well, not only for $\delta_1$ but also for $\delta_3$ and $\delta_5$. 

The experimentally measured $I_3/I_1$ at 15.1$^\circ\,$C and 1$\,$Hz is found to indicate the onset of the transition from linear to non-linear regime at around $\gamma_0 = 0.04 - 0.05$. This can be seen by comparing with the strain sweep data. It shows the maximum of $G^{\prime\prime}$ at $\gamma_0 \approx 0.16$, which is connected with the breaking of the cages. 
This value could be correlated with the raising of $I_5/I_1$ above the `noise' level of $0.1 \cdot I_1/I_1$. The description of $I_3/I_1$ is possible with the formula obtained by Wilhelm ($I_{3/1}(\gamma_0) = A\left[1-(1+(B\gamma_0)^c)^{-1} \right]$) \cite{wilhelm2}. 
For all experimental data the exponent $c$ or the slope of the increase is found to be in the range of $2.2-2.7$. The rescaled simulation results for the harmonics describe the experiment rather closely, whereas the MCT calcuations are found to show a different behaviour. 
All odd harmonics start at low $\gamma_0$ at higher values, whereas for high deformations the intensities coincide with the experimental results. Thus, also the slope of $I_3/I_1$ in this diagram changed to $c \approx 1.7$ instead of the experimental $c = 2.7$.  
These discrepancies between experiment and theory may well be attributable to the slow decay of the correlator to the plateau, as discussed above.    

Whereas the mode-coupling calcuations show a very early transition from the linear to the nonlinear regime 
at $\gamma \approx 0.01$ (see Figs. \ref{exp_f6} to \ref{exp_f8}), for reasons discussed above, the experiments for the frequencies and temperatures investigated show a deviation from linear behaviour at $\gamma \approx 0.04 - 0.05$. 
For the rescaled simulation results of Fig.~\ref{sim_f2}, the deviation starts at $\gamma \approx 0.03$. 
The onset of the yielding process is correlated with the onset of higher harmonics, starting with the third harmonic (see Fig.~\ref{exp_f9}d). 
Although the difference of the onset and intensity of the higher harmonics between theory and experiment is significant, the time signals are only slightly influenced (Fig.~\ref{exp_f2} and \ref{exp_f3}). The FFT is a very sensitive method to analyse the time signals, so very small deviations in the time signal can cause a remarkable difference in the spectra. Increasing the strain amplitude results in more asymmetric time signals with maxima and minima shifted to the left.
    
It was found in experimental, theoretical and simulation strain sweeps that the maximum of $G_1^{\prime\prime}$ lies very close to the crossing point of $G_1^{\prime}(\gamma_0)$ and $G_1^{\prime\prime}(\gamma_0)$. 
For the glassy sample ($\phi=0.65$) this point was located at $\gamma \approx 0.15$.
Beyond this value, the fifth harmonic begins to increase, which can be seen in Fig.~\ref{exp_f9}d for the experimental and rescaled simulation data. 
The theoretical time signals do not show such a sharp transition from the linear to the nonlinear regime. 
Furthermore, the simulation data starts at higher intensity ratios as the experiment; an effect even more pronounced for the theoretical data. 
The schematic model thus predicts a more gradual transition between solid and fluid. 
The onset of the fifth harmonic was also found for micrometer sized particles \cite{Aksel02} to be correlated with the crossing point of $G_1^{\prime}$ and $G_1^{\prime\prime}$. 
The view that the yield strain is indicated by the onset of the fifth harmonic is supported by the results of Petekidis \cite{LeGrand08}. 
For larger deformations, the time signals show a strain softening behaviour, which is typical for shear thinning fluids. The phase shifts of experiment, simulation and theory approach the limiting values of $n\cdot \pi/2$, where $n$ is the index of the harmonic. 
In addition, the harmonics for theory and experiment show the same limiting values lat large $\gamma_0$ for each harmonic (21\% for $I_3/I_1$, 10\% for $I_5/I_1$, 6\% for $I_7/I_1$ and 4\% for $I_9/I_1$). Furthermore the theoretical harmonics exhibit a slight decrease at the highest calculated strains, which is not observed in the experimental data, perhaps due the choice of a too small measurment range. In the simulation the decrease of the intensities is much more pronounced for high deformations, as the intensities do not show a plateau but only a maximum (at 18\% for $I_3/I_1$ and at 8\% for $I_5/I_1$).


\section{Conclusion and outlook}\label{discussion}

We have used a combination of theory, experiment and simulation to investigate the nonlinear stress response 
of dense colloidal dispersions under large amplitude oscillatory shear flow.
The theory employed is a recent extension \cite{pnas} of the well studied F$_{12}^{\dot{\gamma}}$ model \cite{faraday}. 
A key physical mechanism captured by the theory is the yielding  of local particle cages and the subsequent 
onset of nonlinearity. 
In contrast to a related approach presented by Miyazaki \textit{et al.} \cite{miyazaki1,miyazaki3}, the theory employed here makes predictions for the higher harmonic contributions to the stress signal and thus enables the extent of 
the linear response regime to be addressed. 
In order to make contact with experiment, the theory requires only the simultaneously determined parameters from the steady state flow curve and the frequency dependence of the linear moduli to make predictions for the nonlinear viscoelastic response. 
We have thus compared the theory with rheological experiments performed on concentrated suspensions of thermo-sensitive core-shell particles. 

The first nonlinear experiment considered was a strain sweep, performed at three different volume fractions and frequencies. Here we found very good agreement between experiment and theory, although there are small deviations 
concerning the onset of the nonlinear viscoelastic regime. 
The values of $G_1^{\prime}$ and $G_1^{\prime\prime}$ in the linear regime, the crossing point of $G_1^{\prime}$ and $G_1^{\prime\prime}$, maximum of $G_1^{\prime\prime}$ and asymptotic large strain behaviour of $G_1^{\prime}$ and $G_1^{\prime\prime}$ are all well described by the theory. 
Brownian simulation of a two-dimensional system of hard discs enable only a qualitative comparison, but show behaviour broadly consistent with our experimental data, indicating that the yielding process is not strongly dependent upon 
either the material details or dimensionality of the sample. 
Moreover, the simulations do not contain hydrodynamic interactions, suggesting that these are not of great importance for yielding. 
The ratio of the slopes of $G_1^{\prime}$ and $G_1^{\prime\prime}$ for large $\gamma_0$ in the shear molten state 
yields the exponent ratio $\nu'/\nu=2.0\pm0.1$ in theory simulation and experiment. 
The fact that various other materials (of type III in \cite{Hyun02}) display similar asymptotic behaviour may suggest 
a universal mechanism underlying the oscillatory response of shear molten viscoelastic fluids. 

The second type of experiment considered were oscillatory time sweeps. 
In this case the cage yielding is expressed by the deviation of the signal from a sinusoidal form, showing a characteristic asymmetric peak in the yielding regime, which is followed in the shear molten state by a typical 
strain softening semi-circular peak shape. 
The agreement of the time signals obtained from theory, simulation and experiments is good. 
However, small deviations in the time series obtained using the three methods lead to stronger deviations in the parameters of the Fourier transformed time signals. 
This serves to emphasize the fact that FT-rheology is a very sensitive method, sensitive on the logarithmic scale, 
capable of detecting e.g. the fifth harmonic, to an accuracy of less than 1 promille. 
This analysis has shown that the onset of the third harmonic heralds the start of the nonlinear regime and that 
the maximum in $G_1^{\prime\prime}$, which here approximately coincides with the both crossing point of $G_1^{\prime}$ and $G_1^{\prime\prime}$ and the yield strain are correlated with the onset of the fifth harmonic. 
Moreover, the plateau values of the phase shifts at high deformations are found to follow $n\cdot \pi/2$ with 
$n$ beeing the index of each harmonic.

Despite the good overall level of agreement between theory and experiment there remain aspects which could 
be improved. Firstly, the extent (in $\gamma_0$) of the linear response regime is apparently too small within the present theory, with consequences for the variation of the higher harmonic intensities with amplitude. 
As noted, this failing has its origins in the quiescent transient density correlators 
predicted by the $F_{12}$ model, upon which our more recent schematic model is based. 
It would thus be desireable to improve this fundamental aspect of the theory. 
Secondly, the stress response measured in experiment displays a more asymmetric waveform than that predicted by the schematic model. This aspect can potentially be connected to the fact that our model, when applied to calculate the buildup of stress upon the onset of shear, does not generate a stress overshoot. 
While a stress overshoot does occur in approximate solutions of the fully microscopic mode-coupling expressions 
\cite{zausch}, this aspect is lost in making the ansatz (\ref{modulus}) for the shear modulus, as the strain dependent vertex functions appearing in the microscopic theory are replaced by a constant $v_{\sigma}$. 
Empirically reincorporating a strain dependence into the theory via the replacement 
$v_{\sigma}\rightarrow v_{\sigma}(\gamma_0)$, such that the overshoot is recovered, may also lead to an increased waveform asymmetry in the stress response and, thus, better agreement with experiment. 
However, when modifying the schematic model in this fashion, care must be taken not to destroy its existing positive features.

To summarize, it can be concluded that the schmematic mode-coupling model \cite{pnas} can make accurate predictions in the nonlinear viscoelastic regime, based purely on parameters fixed by the steady state stress and linear viscoelastic behaviour. 
This constitutes the first truly time-dependent test (other than step strain) of the schematic model proposed in \cite{pnas}. 
Fourier transform analysis of time series obtained for various strain amplitudes and frequencies provides a wealth 
of experimental information regarding the mechanical response of a material. 
As our present experimental setup enables the investigation of transient flows, we anticipate that the study of such 
flow protocols, in combination with the present results, should enable a complete rheological characterization of our colloidal system.  

\subsection*{Acknowledgements}
Thanks to M. Schwall, J. Grzanna and A. Salah for helping with the programming of the FT-programm.
Financial support was provided by the SFB TR6 and Swiss National Science Foundation (JMB) and 
the SFB 481 (MS).

\end{document}